\documentclass[aps,prb,twocolumn,amsmath,amssymb,amsfonts,floatfix,longbibliography]{revtex4-2}

\usepackage[utf8]{inputenc}
\usepackage[english]{babel}
\usepackage[T1]{fontenc}
\usepackage{float}
\usepackage{graphicx}
\usepackage{dcolumn}
\usepackage{amsmath}
\usepackage{amssymb}
\usepackage{bm}
\usepackage[urlcolor=blue,colorlinks=true,citecolor=red,linkcolor=red,pdfstartview={FitH},bookmarks=false]{hyperref}
\usepackage{xcolor}
\usepackage{listings}
\usepackage{enumitem}
\usepackage{epstopdf}
\usepackage[normalem]{ulem}

\def\vnabla{{\boldsymbol{\nabla}}}

\def\vA{{\bf A}}

\def\vB{{\bf B}}

\def\vzhat{{\hat{\bm{z}}}}

\def\Phiind{{\Phi_{\mathrm{ind}}}}
\def\vj{{\mathbf{j}}}

\def\vm{{\mathbf{m}}}

\def\vvF{{\bf v}_{\mathrm{F}}}
\def\vF{v_{\mathrm{F}}}

\def\vpF{{\bf p}_{\mathrm{F}}}
\def\vpFp{{\bf p}^{\prime}_{\mathrm{F}}}
\def\vR{{\bf R}}

\def\pF{p_{\mathrm{F}}}

\def\EF{E_{\mathrm{F}}}
\def\NF{N_{\mathrm{F}}}

\def\thetaF{{\theta_{\mathrm{F}}}}

\def\muB{\mu_{\mathrm{B}}}
\def\kB{k_{\mathrm{B}}}

\def\Tc{T_{\mathrm{c}}}
\def\TcGamma{T_{\mathrm{c}}^{\Gamma}}
\def\Tcbulk{T_{\mathrm{c}}^{\mathrm{bulk}}}

\def\Omegac{\Omega_{\mathrm{c}}}

\def\Omegabulk{\Omega_0^{\mathrm{bulk}}}

\def\Cbulk{C_c^{\mathrm{bulk}}}

\def\Dxtyt{{\Delta_{d_{x^2-y^2}}}}
\def\Dxy{{\Delta_{d_{xy}}}}

\def\chixtyt{{\chi_{d_{x^2-y^2}}}}
\def\chixy{{\chi_{d_{xy}}}}

\allowdisplaybreaks

\begin{document}

\title{Designing edge currents using mesoscopic patterning in chiral $d$-wave superconductors}

\author{Patric Holmvall}
\email[e-mail:]{patric.holmvall@physics.uu.se}
\affiliation{Department of Physics and Astronomy,
	Uppsala University, Box 516, S-751 20, Uppsala, Sweden}
\author{Annica M. Black-Schaffer}
\affiliation{Department of Physics and Astronomy,
	Uppsala University, Box 516, S-751 20, Uppsala, Sweden}

\date{\today}

\begin{abstract}
Chiral superconductors are topological as characterized by a finite Chern number and chiral edge modes.
Direct fingerprints of chiral superconductivity are thus often taken to be spontaneous edge currents with associated magnetic signatures.
However, a number of recent theoretical studies have shown that the total edge current along semi-infinite edges is greatly reduced or even vanishes in many scenarios for all pairing symmetries except chiral $p$-wave, thus impeding experimental detection.
We demonstrate how mesoscopic finite-sized samples can be designed to give rise to a shape- and size-dependent strong enhancement of the chiral edge currents and their generated orbital magnetic moment and magnetic fields.
In particular, we find that low rotational symmetry systems, such as pentagons and hexagons, give rise to the largest currents, while circular disks also generate large currents but in the opposite direction.
We estimate the resulting magnetic fields to be as large as $0.01\textrm{--}0.5$~mT, with a magnetic moment approaching $\muB/2$ per Cooper pair, where $\muB$ is the Bohr magneton.
The current and magnetic signatures diverge with shrinking system sizes, eventually cut off by finite-size suppression of chiral superconductivity.
We thus also extract the full phase diagram as a function of temperature and system size for different geometries, including competing superconducting orders.
In geometries strongly suppressing only one of the $d$-wave components, we find an additional heat capacity jump, as large as $10\%$ of the bulk normal-superconducting transition, marking the transition between a chiral and a nodal $d$-wave state.
This further acts as an indirect signature of chiral superconductivity, measurable with nanocalorimetry.
Our results are relevant for system sizes on the order of tens to hundreds of coherence lengths, and highlight mesoscopic patterning as a viable route to experimentally identify chiral $d$-wave superconductivity.
\end{abstract}

\maketitle

\section{Introduction}
\label{sec:intro}

Chiral superconductors are a well-studied class of topological matter~\cite{Volovik:1988,Volovik:1992,Read:2000,Volovik:2003,Kallin:2012,Vollhardt:2013,Kallin:2016,Mizushima:2016,Volovik:2019,Volovik:2020}, which spontaneously break time-reversal symmetry in the bulk~\cite{Sigrist:1991} and with topologically protected edge modes associated with a finite Chern number $\nu$~\cite{Volovik:1997,Schnyder:2008,Hasan:2010,Qi:2011,Tanaka:2012,Graf:2013,Black-Schaffer:2014:b}.
The Chern number also directly determines the winding of the superconducting order parameter along the Fermi surface(s) $\Delta(\vpF,\vR) \simeq \Delta(\vR)e^{i\nu\thetaF}$, with the center-of-mass coordinate $\vR$ and Fermi momentum $\vpF$ forming an angle $\thetaF$ on the Fermi surface.

Historically, the focus has mainly been on spin-triplet chiral $p$-wave or $f$-wave superconductivity~\cite{Mackenzie:2003,Kallin:2012,Vollhardt:2013,Kallin:2016,Ran:2019,Li:2019,Suh:2020,Duan:2021,Bae:2021,Hayes:2021,Aoki:2022,Siegl:2024,Xu:2024}, but recently spin-singlet chiral $d$-wave superconductivity has also been proposed in many materials~\cite{
Yamanaka:1998,Takada:2003,Black-Schaffer:2007,Kasahara:2007,Kasahara:2009,Kuroki:2010,Nandkishore:2012,Biswas:2013,Kiesel:2013,Black-Schaffer:2014:b,Fischer:2014,Shibauchi:2014,Gong:2017,Venderbos:2018,Su:2018,Fidrysiak:2018,Xu:2018,Kennes:2018,Liu:2018,Gui:2018,Wu:2019,Hosseinabadi:2019,Ueki:2019,Ueki:2020,Can:2021:a,Can:2021:b,Iguchi:2021,Fischer:2021,Biswas:2021,Ming:2023,Volkov:2025,Pixley:2025:cuprate_review} and even suggested as a platform to realize topological quantum computing~\cite{Nayak:2008,Kitaev:2009,Beenakker:2013,Sato:2017,Mercado:2022,Margalit:2022,Huang:2023,Li:2023,Brosco:2024}.
However, direct experimental verification remains an open problem for all chiral superconductors~\cite{Bjornsson:2005,Kirtley:2007,Hicks:2010,Iguchi:2024}. Most efforts have centered around 
trying to detect the chiral edge modes, specifically their associated charge currents and magnetic fields~\cite{Anderson:1961,Volovik:1975,Leggett:1975,Ishikawa:1977,Cross:1977,Leggett:1978,McClure:1979,Sigrist:1990,Sauls:1994,Kita:1998,Matsumoto:1999,Matsumoto:1999_errata,Furusaki:2001,Stone:2004,Stone:2008,Sauls:2011,Byun:2018,Holmvall:2023:enhanced,Pathak:2024}, although recent theoretical proposals also include signatures of Abrikosov vortices~\cite{Yoshida:2024}, coreless vortices~\cite{Holmvall:2023:cv,Holmvall:2023:robust}, and their analogous skyrmionic chains~\cite{Cadorim:2024}. 

\begin{figure}[t!]
	\includegraphics[width=\columnwidth]{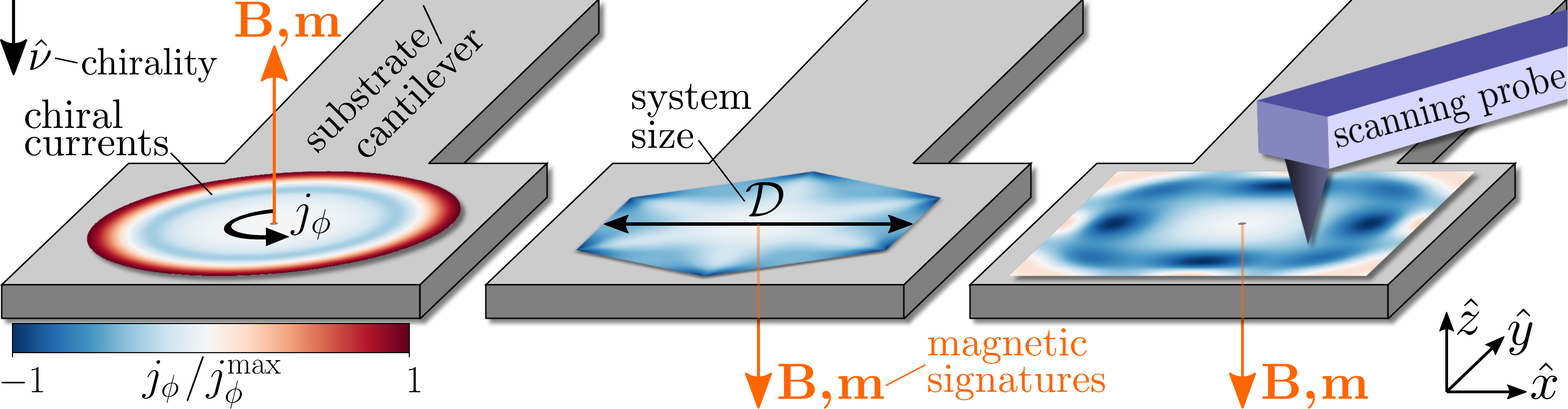}
	\caption{Sketch of thin chiral $d$-wave superconductors (SC) with negative chirality ($\hat{\bm \nu}=-\hat{\bm z}$) with system size $\mathcal{D}$ tens of $\xi_0$ and different mesoscopic shapes. Heatmap indicates the chiral charge-current density $j_\phi(\vR)$, inducing an orbital magnetic moment $\vm = \pm m_z \hat{\bm z}$ and magnetic field $\vB = \pm B_z\hat{\bm z}$ with directions depending on the SC shape, measured with e.g.~a scanning probe (purple) or cantilever setup (gray). }
	\label{fig:sketch_setup}
\end{figure}

Chiral edge modes carry a particle current, which should generate a spontaneous edge charge current with associated magnetic fields~\cite{Sigrist:1990} and orbital magnetic moments~\cite{Spaldin:2008,SuperConga:2023}, see Fig.~\ref{fig:sketch_setup}.
However, in contrast to Chern insulators, the edge charge currents in chiral superconductors are not topologically protected due to the non-conserved charge in superconductors~\cite{Volovik:1988,Furusaki:2001,Stone:2004,Black-Schaffer:2012}, but may instead depend on various system details, including edge geometry, boundary conditions, impurities, Fermi surface shape, and band effects~\cite{Ashby:2009,Sauls:2011,Bouhon:2014,Etter:2014,Huang:2014,Huang:2015,Tada:2015,Volovik:2015:b,Ojanen:2016,Suzuki:2016,Suzuki:2017,Goryo:2017,Wang:2018,Tada:2018,Etter:2018,Nie:2020,Suzuki:2022,Suzuki:2023,Holmvall:2023:enhanced,Suzuki:2024:arxiv,Higashitani:2024}.
In fact, it has been shown theoretically that the total, or net, edge charge current may be much lower in chiral $d$-wave superconductors compared to chiral $p$-wave superconductors~\cite{Huang:2014,Wang:2018}, even completely vanishing in certain limits, regardless of the magnetic screening~\cite{Huang:2014,Huang:2015,Tada:2015,Volovik:2015:b}.
The vanishing net charge current has been explained in terms of canceling contributions between the chiral edge modes and the backflow response of the superconducting condensate~\cite{Sauls:2011,Nie:2020}.
However, the topologically unprotected nature of the edge charge currents also means that control of the environment can be used advantageously to engineer and even greatly enhance the net charge current.
For instance, we recently showed~\cite{Holmvall:2023:enhanced} that mesoscopic finite-size effects may cause a large net charge current and magnetic signatures measurable with e.g.~scanning probes and cantilever setups~\cite{Bolle:1999,Khotkevych:2008,Bleszynski-Jayich:2009,Bert:2011,Jang:2011,Vasyukov:2013,Curran:2014,Kirtley:2016,BishopVanHorn:2019,Uri:2020,Kim_Schwenk:2021,Persky:2022}, see Fig.~\ref{fig:sketch_setup}.

In this work, we show that the spontaneous edge charge currents and the associated magnetic signatures in mesoscopic superconductors do not just have a size dependence as previously established~\cite{Holmvall:2023:enhanced}, but most importantly also very strong shape dependencies, as schematically illustrated in Fig.~\ref{fig:sketch_setup}.
Specifically, we demonstrate that a strong current enhancement can be engineered, yielding magnetic fields on the order of $0.01\textrm{--}0.5$~mT and a large orbital magnetic moment of $\muB/2$ per Cooper pair, where $\muB$ is the Bohr magneton.

We find that superconductors shaped as pentagons, hexagons, or circular disks, are particularly promising with experimentally measurable signatures present in systems with size $\mathcal{D}$ of the order of tens of the superconducting coherence length $\xi_0$, thus typically tens to hundreds of nanometers.
In contrast, square and triangle geometries hosts by far the smallest currents and magnetic signatures.
More specifically, we consider two-dimensional (2D) chiral $d$-wave superconductors of different shapes and as a function of temperature and system size, see Fig.~\ref{fig:sketch_setup}, focusing on both regular polygons and circular disks with feasible experimental realizations~\cite{Geim:1997,Chibotaru:2000,Kanda:2004,Grigorieva:2006,Kokubo:2010,Cren:2011,Gustafsson:2013,Timmermans:2016,Curran:2023}.
Generally we find that in finite-sized systems both the charge-current magnitude and direction depend strongly on the system shape.
Specifically, polygons with low rotation symmetry (e.g.~pentagons, hexagons) have a large net charge current parallel to the chirality, which for increasing rotation symmetry evolves smoothly towards an also large net charge current in the circular disk, but now instead antiparallel to the chirality.
System shapes in-between therefore often generate diminishing currents.
We relate these results to edge-edge interference, as well as hybridization between edge modes~\cite{Sauls:2011,Wu:2023}, which are always present in finite-size systems.

Overall, we find that the magnitude of the charge current and magnetic signatures diverge with decreasing system size for all systems shapes, explained by increasingly stronger finite-size effects.
Eventually, the divergence is cut-off below $\mathcal{D} \lesssim 10\text{--}20\xi_0$ because of an overall finite-size suppression of the superconducting order parameter.
The order parameter suppression further induces a competition between different chiral and nodal superconducting states, which we fully quantify via the ground-state phase diagram as a function of temperature and system size for several geometries.
We find that each transition in the phase diagram is of second order with an associated jump in the heat capacity $C(T)$.
Interestingly, in geometries that break the degeneracy between the underlying nodal components, the onset of chiral superconductivity occurs via an intermediate nodal superconducting state.
This transition is marked by an additional jump as large as $10\%$ of the bulk normal-superconducting transition and a crossover from polynomial to exponential $C(T)$, both of which could be measured with nanocalorimetry~\cite{Tagliati:2012,Willa:2017,Feng:2019}.
When instead the system size increases towards the limit $\mathcal{D}\to\infty$, all shapes eventually show vanishing net charge current and magnetic signatures, fully consistent with previous studies~\cite{Huang:2014,Huang:2015,Tada:2015,Volovik:2015:b,Nie:2020}.
However, due to a finite edge curvature we find a slower asymptotic decay in disk-shaped systems, with finite net charge currents even up to $\mathcal{D} \approx 200\xi_0$.
Taken together, our results highlight mesoscopic patterning as a viable route to experimentally engineer and enhance the signatures of chiral superconductivity.

The remainder of this work is organized as follows.
We describe our theoretical methods and models in Sec.~\ref{sec:methods}.
In Sec.~\ref{sec:results:currents} we investigate how the charge-current density depends on system size and shape, first by studying its spatial dependence and then by focusing on the net, or integrated, current and its associated magnetic signatures.
In Sec.~\ref{sec:results:phase_diagrams} we investigate how confinement also induces a competition between different pairing symmetries by extracting the ground-state phase diagram as a function of temperature and system size, for different system shapes. 
Finally, we conclude our work in Sec.~\ref{sec:conclusions}.

\section{Methods and models}
\label{sec:methods}
In this section we provide a brief overview of the theoretical formalism, order parameter pairing symmetry, model system, and numerical details used in this work.

\subsection{Quasiclassical theory}
\label{sec:methods:quasiclassics}
We are interested in the emergent properties of a chiral $d$-wave superconductor under the influence of mesoscopic confinement, together with the asymptotic behavior as the system size grows towards the bulk regime. To accurately capture the full behavior, including edge charge currents, we need to implement full self-consistency for both the superconducting order parameter $\Delta$ and vector potential $\vA$.
This leads to challenging calculations, which we tackle using the well-established quasiclassical theory of superconductivity~\cite{Eilenberger:1968,Larkin:1969,Eliashberg:1971,Serene:1983,Shelankov:1985,Nagato:1993,Eschrig:1994,Schopohl:1995,Schopohl:1998,Eschrig:1999,Belzig:1999,Kopnin:2009,Eschrig:2009,Grein:2013,Eschrig:2015}.
This subsection gives a brief background to the theory, see also Refs.~\cite{Holmvall:2023:enhanced,Holmvall:2023:robust,SuperConga:2023}.

The quasiclassical theory of superconductivity is a controlled expansion in parameters that are typically small in many metals, such as $|\Delta|/\EF$ and $\hbar/(\pF\xi_0)$, where $\EF$ is the Fermi energy, $\hbar$ the reduced Planck constant, $\pF$ the Fermi momentum on the Fermi surface, $\xi_0=\hbar\vF/(2\pi\kB\Tc)$ the effective superconducting coherence length with Fermi velocity $\vF$, Boltzmann's constant $\kB$, and critical temperature $\Tc$.
In particular, we use the formulation of Eilenberger~\cite{Eilenberger:1968}, where the central object is the quasiclassical propagator $\hat{g}(\vpF, \vR; z)$, with center-of-mass coordinate $\vR$ and complex energy $z$.
We express this propagator in Nambu space via
\begin{align}
\label{eq:model:green_function}
\hat{g}(\vpF, \vR; z) = 
\begin{pmatrix}
    g(\vpF, \vR; z) & f(\vpF, \vR; z)\\
    -\tilde{f}(\vpF, \vR; z) & \tilde{g}(\vpF, \vR; z)
\end{pmatrix},
\end{align}
where $g(\vpF, \vR; z)$ and $f(\vpF, \vR; z)$ are the quasiparticle and anomalous pair propagators in spin space, respectively, with ``tilde'' denoting particle-hole conjugation
$\tilde{\alpha}(\vpF, \vR; z) = \alpha^*(-\vpF, \vR; -z^*)$.
In the following, we briefly drop the arguments for brevity.
The propagators are obtained from the Eilenberger equation~\cite{Eilenberger:1968}
\begin{align}
    \label{eq:model:eilenberger}
    i\hbar\vvF\cdot\boldsymbol{\nabla}\hat{g} + \left[z\hat{\tau}_3 - \hat{h},\hat{g}\right] = 0,
\end{align}
which is a transport-like equation describing quasiparticle and pair propagation along the Fermi velocity $\vvF$, with normalization condition $\hat{g}^2 = -\pi^2\hat{\tau}_0$, Pauli-spin matrices $\hat{\tau}_i$ in Nambu space. The self-energies $\hat{h}$ are separated into diagonal ($\hat{\Sigma}$) and off-diagonal ($\hat{\Delta}$) parts
\begin{align}
    \label{eq:model:self_energy}
    \hat{h} = \hat{\Sigma} + \hat{\Delta} = \begin{pmatrix}
    \Sigma & \Delta \\
    \tilde{\Delta} & \tilde{\Sigma}
    \end{pmatrix}.
\end{align}
In the following, we first describe the diagonal and then the off-diagonal self-energy terms.

While we do not consider any externally applied magnetic fields, chiral superconductivity leads to finite charge-current densities $\vj(\vR)$ e.g.~at the system edges~\cite{Sauls:2011}, which together with finite London-penetration depth $\lambda_0$ generates a finite gauge field via Amp{\`e}re's law
\begin{align}
    \label{eq:model:ampere}
    \vnabla\times\vB(\vR) = \vnabla \times \vnabla \times \vA(\vR) = \frac{4\pi}{c}\vj(\vR).
\end{align}
This generates a diagonal self-energy
\begin{equation}
    \label{eq:diagonal_self_energy}
    \hat{\Sigma} = -\frac{e}{c}\vvF(\vpF)\cdot\vA(\vR)\hat{\tau}_3,
\end{equation}
which requires a fully self-consistent solution for the gauge field $\vA$, here with elementary charge $e=-|e|$ and speed of light $c$. To achieve this self-consistency, we compute the charge-current density via
\begin{align}
    \label{eq:model:current_density}
    \vj(\vR) = 2e \NF \kB T\sum_{n}^{|\varepsilon_n|<\Omegac} \left\langle \vvF(\vpF)\, g(\vpF,\vR;\varepsilon_n) \right\rangle_{\vpF},
\end{align}
which we express in units of $j_0 \equiv \hbar|e|\vF^2\NF/\xi_0 = 2\pi|e|\kB\Tc\NF\vF$, with normal-state density of states $\NF$ (per spin).
The sum in Eq.~(\ref{eq:model:current_density}) runs over the Matsubara energies $i\varepsilon_n = i\pi\kB T(2n+1)$ with temperature $T$ and integer $n$, which we calculate using the efficient ``Ozaki summation''~\cite{Ozaki:2007} based on the Matsubara technique~\cite{Matsubara:1955,Bruus:2004,Rammer:2007,Kopnin:2009,Mahan:2013}, where we also use a standard approach to eliminate the cutoff $\Omegac$ in favor of $\Tc$~\cite{Grein:2013}.
The angle brackets denote Fermi surface integration~\cite{Graf:1993}
\begin{align}
    \label{eq:model:fs_average}
    \left\langle \dots \right\rangle_{\vpF} = \frac{1}{\NF} \oint \frac{\mathrm{d} p_\mathrm{F}}{(2\pi\hbar)^2|\vvF(\vpF)|} (\dots).
\end{align}
The charge-current density also gives rise to an orbital magnetic moment (OMM) $\vm = m_z\vzhat$ (per 2D layer with area $\mathcal{A}$)~\cite{Spaldin:2008,SuperConga:2023}
\begin{align}
    \label{eq:model:magnetic_moment}
    \frac{\vm}{m_0} \equiv 2\int_{\mathcal{A}} \frac{d\vR}{\mathcal{A}} \frac{\vR}{\xi_0} \times \frac{\vj(\vR)}{j_0},
\end{align}
with natural units $m_0 \equiv \muB (N/2)$, Bohr magneton $\muB = \hbar|e|/2m^*$, $N/2$ Cooper pairs, and effective quasiparticle mass $m^*$ defined via $\vpF = m^*\vvF$.
We additionally extract the total induced flux through each layer
\begin{align}
    \label{eq:model:total_flux}
    \Phiind = \int_{\mathcal{A}} d\vR \cdot \vB(\vR),
\end{align}
with flux quantum $\Phi_0 \equiv hc/2|e|$ and Planck constant $h$.

The off-diagonal self-energy $\Delta$ is the mean-field superconducting order parameter, which we obtain fully self-consistently from the superconducting gap equation
\begin{align}
    \label{eq:model:gap_equation}
    \Delta(\vpF,\vR) = \NF \kB  T\sum_n^{|\varepsilon_n|<\Omegac} \big\langle
                  V(\vpF,\vpFp)\,f(\vpFp,\vR;\varepsilon_n)
                   \big\rangle_{\vpFp},
\end{align}
where we decompose the effective pairing interaction $V(\vpF,\vpFp)$ into the even-parity spin-singlet symmetry channels (i.e.~even-parity irreducible representations of the crystallographic point group)~\cite{Yip:1993},
\begin{align}
    \label{eq:model:pairing_interaction}
    V(\vpF,\vpFp)=\sum_{\Gamma} V_\Gamma \eta_\Gamma(\vpF)\eta^\dagger_\Gamma(\vpFp).
\end{align}
Here, $\Gamma$ labels the irreducible representation, $V_\Gamma$ is the pairing strength of the respective symmetry channel, and $\eta_\Gamma(\vpF)$ is the basis function encoding the pairing symmetry on the Fermi surface.
The total superconducting order parameter $\Delta(\vpF,\vR)$ can be decomposed as
\begin{align}
    \label{eq:model:order_parmeter:irreducible_representation}
    \Delta(\vpF,\vR) = \sum_\Gamma|\Delta_\Gamma(\vR)|e^{i\chi_\Gamma(\vR)}\eta_\Gamma(\vpF),
\end{align}
where each symmetry channel is associated with an order parameter component $\Delta_\Gamma(\vpF,\vR)$ with amplitude $|\Delta_\Gamma(\vR)|$ and phase $\chi_\Gamma(\vR)$, as well as bulk critical temperature $\TcGamma$.
Our only assumption about the superconducting order parameter is to have equal $\TcGamma$ for the two $d$-wave channels, as further described in Sec.~\ref{sec:methods:order_parameter}.

To compare different solutions, we compute the free energy $\Omega$ as the difference between the superconducting (S) and normal (N) states, $\Omega = \Omega_{\rm S} - \Omega_{\rm N}$, from the Luttinger-Ward potential~\cite{Luttinger:1960,Serene:1983,Thuneberg:1984,Vorontsov:2003,Virtanen:2020},
\begin{align}
    \nonumber
    \Omega = \int d\vR \Bigg\{\frac{|\vB(\vR)|^2}{8\pi} + \NF\sum_{\Gamma}\left\langle|\Delta(\vpF,\vR)|^2\right\rangle_{\vpF}\ln\frac{\Tc}{\TcGamma}\\
    \label{eq:free_energy}
    + \pi\NF\kB T\sum_{n}\left\langle\frac{|\Delta(\vpF,\vR)|^2}{|\varepsilon_n|} - \mathcal{I}(\vR)\right\rangle_{\vpF}
    \Bigg\},
\end{align}
where we use the Eilenberger form of the last term~\cite{Eilenberger:1968}
\begin{align}
\label{eq:free_energy:kernel:eilenberger}
    \mathcal{I}(\vR) = \frac{\Tilde{\Delta}(\vpF,\vR)f(\vpF,\vR;\varepsilon_n)+\Delta(\vpF,\vR)\Tilde{f}(\vpF,\vR;\varepsilon_n)}{\pi + ig(\vpF,\vR;\varepsilon_n)}.
\end{align}
The corresponding heat capacity difference $C = C_{\rm S} - C_{\rm N}$ is obtained via the thermodynamic definition
\begin{equation}
    C = -T\frac{\partial^2\Omega}{\partial T^2}.
\end{equation}

\subsection{Chiral and nodal pairing symmetries}
\label{sec:methods:order_parameter}

\begin{figure}[t!]
	\includegraphics[width=\columnwidth]{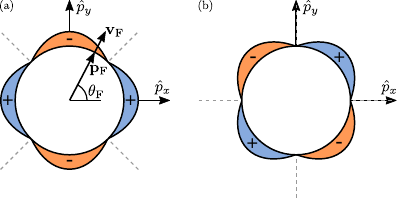}
	\caption{Schematic illustration of cylindrical Fermi surfaces in the 2D plane with nodal $d$-wave superconducting gaps (colored) corresponding to the basis functions (a) $\eta_{d_{x^2-y^2}}(\vpF) = \sqrt{2}\cos\left(2\thetaF\right)$ and (b) $\eta_{d_{xy}}(\vpF) = \sqrt{2}\sin\left(2\thetaF\right)$. The crystal $ab$-axes coincide with the principal momentum directions $\hat{p}_x$ and $\hat{p}_y$. Gray dashes are guides to the eye marking the nodal gap directions.}
	\label{fig:basis_functions}
\end{figure}

In this work we consider chiral $d$-wave superconductivity in 2D, which requires attraction in the two pairing channels $\Gamma \in \{d_{x^2-y^2}, d_{xy}\}$ with $\eta_{d_{x^2-y^2}}(\vpF) = \sqrt{2}\cos\left(2\thetaF\right)$ and $\eta_{d_{xy}}(\vpF) = \sqrt{2}\sin\left(2\thetaF\right)$, see Fig.~\ref{fig:basis_functions} for a schematic illustration.
The resulting total order parameter can then be written as
\begin{align}
    \label{eq:model:order_parameter:dwaves}
    \Delta(\vpF,\vR) & = \Delta_{d_{x^2-y^2}}(\vpF,\vR) + \Delta_{d_{xy}}(\vpF,\vR),
\end{align}
where each component has its amplitude and phase as in Eq.~(\ref{eq:model:order_parmeter:irreducible_representation}).
The only assumption we make is to assume the same pairing strength in both $d$-wave channels, hence the same $\TcGamma$. The same $\TcGamma$ for both $d$-wave solutions
is guaranteed by the lattice symmetry for all lattices with three- or six-fold rotation symmetry, which occurs in many materials suggested to be chiral $d$-wave superconductors \cite{Black-Schaffer:2007,Black-Schaffer:2014:b,Nandkishore:2012, Venderbos:2018,Su:2018,Fidrysiak:2018,Xu:2018,Kennes:2018,Liu:2018,Gui:2018,Wu:2019,Fischer:2021,Biswas:2013,Fischer:2014,Ueki:2019,Ueki:2020,Takada:2003,Kiesel:2013,Yamanaka:1998,Kuroki:2010,Ming:2023}, or may be engineered by a $45^\circ$ twist between two layers with $d$-wave superconducting symmetry \cite{Can:2021:a,Can:2021:b}.
We then let the order parameter, amplitude and phase, of both $d$-wave components evolve completely independently in our self-consistency calculation of the ground state. 
This means that there is always a competition
between different $d$-wave superconducting solutions, as well as with the normal state.
Any superconducting state with a relative phase difference $\chi_{\rm rel}\equiv \chixtyt-\chixy \in \{0,\pi\}$ gives a $d$-wave like order parameter with nodal lines along some direction in the 2D plane.
This state preserves time-reversal symmetry, has no finite Chern number, no chiral edge states, and no spontaneous charge currents.
We note, however, that for certain edge terminations, different $d$-wave components may become completely suppressed by surface pair-breaking, causing non-dispersive and flat bands of Andreev bound states to emerge~\cite{Hu:1994,Jian:1994,Lofwander:2001,Vorontsov:2018}.
In contrast, for any other relative phase difference, the order parameter will instead be fully gapped.
In fact, in all our calculations we only find $\chi_{\rm rel} = 0$ in a few special scenarios, which we call the nodal phase, and in all other cases we find $\chi_{\rm rel} = \pm \pi/2$.
The state with $\chi_{\rm rel} = \pm \pi/2$ is the fully symmetric chiral state, which breaks time-reversal symmetry in the bulk~\cite{Sigrist:1991} (the positive and negative solutions are related via the time reversal operation).
This chiral state is an eigenstate of the orbital angular momentum (OAM) operator $\hat{L}_z^{\rm orb}\Delta(\vR,\vpF) = l_z^{\rm orb}\Delta(\vR,\vpF)$ where $l_z^{\rm orb}=\nu\hbar$ is the OAM carried by each Cooper pair~\cite{Sauls:1994}, with Chern number $\nu =\pm2$ for a chiral $d$-wave superconductor in 2D~\cite{Volovik:1997,Schnyder:2008,Hasan:2010,Qi:2011,Tanaka:2012,Graf:2013,Black-Schaffer:2014:b}.
Edges of the system give rise to $|\nu|$ number of topologically protected chiral edge modes, due to the bulk-boundary correspondence~\cite{Volovik:1997,Schnyder:2008,Hasan:2010,Qi:2011,Tanaka:2012,Graf:2013,Black-Schaffer:2014:b}.
These chiral edge modes are dispersive and generate a spontaneous charge-current density $\vj(\vR)$~\cite{Sauls:2011}, which in turn generate spontaneous magnetic fields $\vB(\vR)$ via Eq.~(\ref{eq:model:ampere}) and spontaneous OMM $m_z$ via Eq.~(\ref{eq:model:magnetic_moment}), see Fig.~\ref{fig:sketch_setup}.

\subsection{Model}
\label{sec:methods:model}
In this work, we focus on the influence of 2D confinement on a chiral $d$-wave superconducting state with negative chirality $\hat{\nu}=-\hat{z}$ in the bulk or interior of the system.
We note that similar mesoscopic confinement has previously been shown to give rise to many fascinating superconducting and superfluid
phenomena~\cite{Geim:1997,Schweigert:1998,Chibotaru:2000,Kanda:2004,Grigorieva:2006,Vorontsov:2009,Kokubo:2010,Cren:2011,Zhang:2012,Zhang:2013,Gustafsson:2013,Timmermans:2016,Wiman:2018,Vorontsov:2018,Levitin:2019,Shook:2020,Levitin:2020:comment,Shook:2020:reply,Choi:2021,Regan:2021,Varga:2022,Yapa:2022,Wu:2023,Pleiner:2023,Sun:2023}.
We further assume weak-coupling superconductivity in equilibrium with full spin degeneracy.
For simplicity, we additionally assume a cylindrically symmetric Fermi surface~\cite{Graf:1993} and specular superconductor-vacuum surfaces~\cite{SuperConga:2023}, leaving effects of normal-state anisotropy and disorder as an interesting future extension of our work.
We study the ground state as a function of both the shape and size $\mathcal{D}$ of the superconductor, from the small mesoscopic limit ($\mathcal{D} \sim \xi_0$) to the macroscopic semi-infinite limit ($\mathcal{D} \gg \xi_0$), at different temperatures $T$ and also with a finite penetration depth $\lambda_0 \equiv \sqrt{c^2/\left(4\pi e^2 \vF^2\NF\right)}$.
We thus include effects of spontaneous Meissner screening of the chiral currents caused by finite $\lambda_0$ but we generally find negligible influence of such screening, except when $\xi_0 \sim \lambda_0 \ll \mathcal{D}$~\cite{Holmvall:2023:enhanced}.
However, since most thin or unconventional superconductors are instead extreme Type-II ($\xi_0 \ll \lambda_0$)~\cite{Kogan:2014:a,Kogan:2014:b,Ooi:2021,Prozorov:2022}, we for simplicity focus on $\lambda_0\to\infty$ in most of our results, except when computing the spontaneously induced flux $\Phiind$.

\subsection{Numerics}
\label{sec:methods:numerics}
To solve the Eilenberger equation~\eqref{eq:model:eilenberger} we use the efficient and numerically stable Riccati formalism~\cite{Nagato:1993,Schopohl:1995,Schopohl:1998,Eschrig:2000}, implemented in the open-source framework SuperConga~\cite{SuperConga:2023}.
This user-friendly~\cite{SuperConga:documentation} framework is free to download~\cite{SuperConga:repository} and trivializes the setup for differently shaped superconductors, while still providing accurate self-consistent solutions also for numerically extremely challenging sizes.
Specifically, SuperConga solves Eqs.~(\ref{eq:model:eilenberger}), (\ref{eq:model:ampere}) and (\ref{eq:model:gap_equation}) self-consistently in an iterative process until the global error $\epsilon_{\mathrm{G}}$ of quantity $O_i$ at iteration number $i$ is $\epsilon_{\mathrm{G}} = \left \| O_i-O_{i-1} \right \|_2/\left \|O_{i-1} \right \|_2 < \epsilon_{\mathrm{tol}}$, for $O \in \{\Delta(\vpF,\vR),\vA(\vR),\vj(\vR),\Omega\}$ with tolerance $\epsilon_{\mathrm{tol}}$.
For accurate solutions, we set the condition for self-consistency at $\epsilon_{\mathrm{tol}} = 10^{-7}$, which we reach using a combination of the sophisticated convergence accelerators CongAcc~\cite{SuperConga:2023} and Barzilai-Borwein~\cite{BarzilaiBorwein:1988}.
We use the energy cutoff $\Omegac \gtrsim 100 \kB\Tc$, a discrete spatial resolution of $20$ points per coherence length, and a discrete angular resolution of $256$ points on the Fermi surface.
These values were chosen such that we do not notice any difference with finer resolution.
We always keep the same spatial resolution when changing the system size, and keep all parameters fixed during the process of converging towards the self-consistent solution.
For further details on implementation and numerics, see Ref.~\cite{SuperConga:2023}.

\section{Shape and Size Enhanced charge currents and magnetic signatures}
\label{sec:results:currents}
In this section, we present our main results when it comes to the current and magnetic signatures of differently shaped and sized mesoscopic chiral $d$-wave superconductors.
A vanishing net current has repeatedly been reported in generic treatments of chiral $d$-wave superconductors~\cite{Huang:2014,Huang:2015,Tada:2015,Volovik:2015:b}.
The vanishing current can be understood in terms of effective cancellation of contributions between dispersive chiral edge modes and condensate backflow~\cite{Sauls:2011,Nie:2020}.
Specifically, these contributions lead to a charge-current density that changes sign at a very small distance $\sim \xi_0$ from the edge, see Fig.~\ref{fig:current_density}(a) for an illustration, where the near-edge (positive) and the far-edge (negative) portions of the charge current average to zero when spatially integrated.
Here we demonstrate how the chiral edge currents can be engineered via mesoscopic patterning to instead be significantly enhanced, thus aiding experimental efforts to detect chiral superconductivity.
In all systems with large net currents, we also find accompanied notable magnetic signatures, within experimental reach.

\subsection{Charge-current density}
\label{sec:results:currents:current_density}
\begin{figure}[t!]
	\includegraphics[width=\columnwidth]{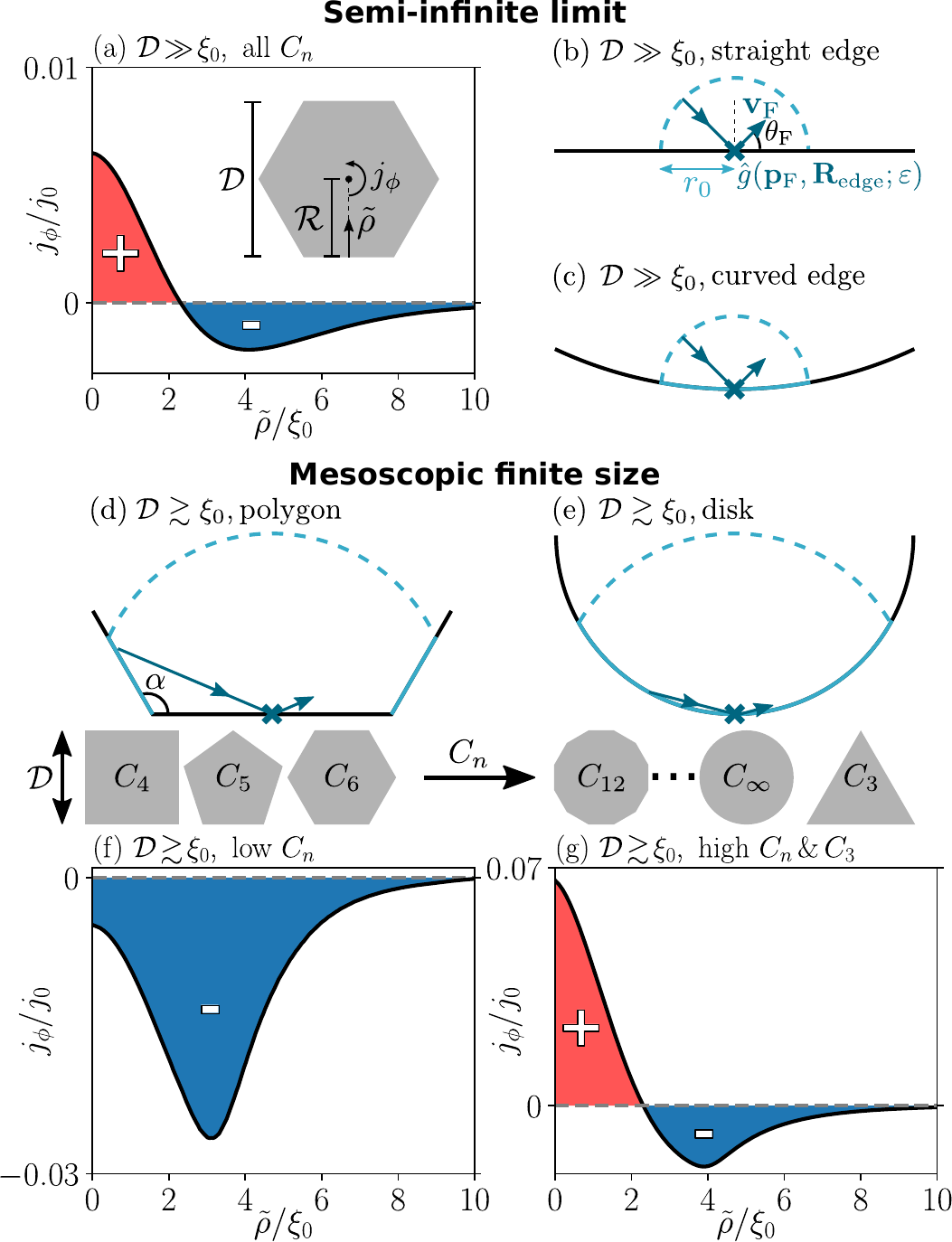}
	\caption{(a) Typical spatial dependence of the charge-current density $j_\phi$ as a function of distance $\tilde{\rho}$ from the edge in the semi-infinite limit $\mathcal{D} \gg \xi_0$ for all shapes at low temperature $T=0.1\Tc$ for a negative chirality ($\hat{\nu} = -\hat{z}$)  $d$-wave superconductor, with natural units $j_0 \equiv \hbar|e|\vF^2\NF/\xi_0 = 2\pi|e|\kB\Tc\NF\vF$. Here, $(\rho,\phi, z)$ are polar coordinates with $\tilde{\rho} \equiv \mathcal{R} -\rho$ being the radial coordinate measured from the edge towards the system center, with apothem $\mathcal{R} \sim \mathcal{D}/2$, see inset.
    (b),(c) Schematic illustrations of region of size $r_0 \sim 2\text{--}10\xi_0$ with finite contributions (blue) to the surface propagator (cross) at a straight and curved edges (solid black), respectively. The contributions are mediated via ballistic trajectories along the Fermi velocity $\vvF$ (arrows) and are separated into bulk-edge (dashed blue) and edge-edge (solid blue) contributions.
    (d) Same as (b,c) but in a small finite-sized system shaped like a regular polygon with $n$ edges and internal angle $\alpha = \pi(n-2)/n$, and (e) in a disk.
    (f),(g) Same as (a) but in finite-sized systems $\mathcal{D} \gtrsim \xi_0$ with low and high rotation symmetries $C_n$, respectively. The triangle is an outlier belonging to the latter, see main text.}
	\label{fig:current_density}
\end{figure}

We first present how the charge-current density spatially varies as a function of distance from the edge across different system shapes and sizes, starting with systems approaching the semi-infinite limit $\mathcal{D} \gg \xi_0$, before uncovering how this behavior completely changes depending on system shape in finite-sized systems $\mathcal{D} \gtrsim \xi_0$.
Here, $\mathcal{D}$ denotes the full distance across the system, while $\mathcal{R} \sim \mathcal{D}/2$ is the edge-center distance, i.e.~the radius (apothem) in a disk (polygon), see Fig.~\ref{fig:current_density}(a).
We focus on systems with negative chirality $\hat{\nu} = -\hat{z}$ and at low temperature $T=0.1\Tc$.
For negative chirality, the edge mode dispersion is such that it naively should carry a positive charge current.

For sufficiently large system size $\mathcal{D}\gg\xi_0$ we find that for all system shapes the charge-current density takes the typical form given in Fig.~\ref{fig:current_density}(a).
Integrating to find the net current, this reproduces earlier results of a vanishing net current in the semi-infinite limit~\cite{Huang:2014,Huang:2015,Tada:2015,Volovik:2015:b}.
Thus, despite the guaranteed existence of chiral edge modes due to the bulk-boundary correspondence~\cite{Volovik:1997,Schnyder:2008,Hasan:2010,Qi:2011,Tanaka:2012,Graf:2013,Black-Schaffer:2014:b}, there is no net charge current. 

To understand how the zero net current in the semi-infinite size limit develops into finite currents for finite-sized systems, we first need to understand how the charge-current density $\vj(\vR)$ in Eq.~(\ref{eq:model:current_density}) depends on the local environment.
We schematically illustrate this dependence based on the quasiclassical theory of superconductivity, where the dependence is directly encoded in the surface propagator $\hat{g}(\vpF,\vR;\varepsilon)$ via Eq.~(\ref{eq:model:eilenberger}).
For a clean system, this surface propagators has a spatial dependence typically described by exponentially decaying solutions along straight ballistic trajectories along $\vvF$~\cite{Sauls:2011,Wu:2023,Holmvall:2023:enhanced} as described by the Eilenberger equation (\ref{eq:model:eilenberger}) and illustrated by blue lines with arrows in Fig.~\ref{fig:current_density}(b).
This spatial dependence leads to an effective correlation length of the surface propagators, marked by blue dashes in Fig.~\ref{fig:current_density}(b) and denoted by $r_0 \sim 2\text{--}10\xi_0$, with exact values depending on e.g.~temperature~\cite{Holmvall:2023:enhanced}.
This emergent length scale describes the healing length of the superconducting order parameter away from the edge, and also sets the decay length towards the bulk for the chiral edge modes, as well as the currents they generate~\cite{Holmvall:2023:enhanced}.
For a semi-infinite system, the surface propagators only contain edge-bulk interactions, as the edge cannot interact with itself at a planar surface, see Fig.~\ref{fig:current_density}(b).
However, for finite edge curvatures and finite $\mathcal{D}$, edge-edge contributions can also appear in the surface propagators, see colored portion of the edge in Fig.~\ref{fig:current_density}(c).
Specifically, the curvature allows for separate points along the system edge with different surface normals to be connected by straight ballistic quasiparticle trajectories. 
Such points, separated by distances $\lesssim r_0$, lead to additional edge-edge contributions in the surface propagators~\cite{Sauls:2011,Wu:2023}, which then also modify the charge-current density through Eq.~(\ref{eq:model:current_density}), leading to shape effects for the current in finite-size systems.
Since these are edge-edge contributions in propagators, they can also be viewed as an interference effect between different parts of the sample edge.
A direct consequence of this is that in the disk, the charge-current density approaches the result of the semi-infinite limit more slowly than in other shapes, see Appendix~\ref{app:current_density} for an explicit comparison.

We turn next to more directly analyzing finite-sized systems $\mathcal{D} \gtrsim \xi_0$, such as those schematically illustrated in Figs.~\ref{fig:current_density}(d) and \ref{fig:current_density}(e).
Here, edge portions separated by a distance $\lesssim r_0$ generate significant edge-edge interactions between the chiral edge modes.
In particular, in systems with low rotation symmetry the edge-edge interactions, or interference, occurs between edges that have a more perpendicular relative orientation, see Fig.~\ref{fig:current_density}(d), while for systems with high rotation symmetry the relative orientation is more parallel, see Fig.~\ref{fig:current_density}(e). 
We consistently find that the interactions between edges with a more perpendicular (parallel) relative orientation generate an overall destructive (constructive) interference for the near-edge charge-current density, resulting in negative (positive) charge-current density in this region.
With destructive (constructive) we here mean interference that destroy (enhance) the naively positive edge mode current generated by its innate dispersion.
As a consequence, we generally find an overall negative charge-current density in systems with low rotation symmetry, such as pentagons and hexagons, as illustrated in Fig.~\ref{fig:current_density}(f), which evolves smoothly with increased rotation symmetry $C_n$ towards an overall positive charge-current density in the disk limit $n\to\infty$ see Fig.~\ref{fig:current_density}(g).
In-between we find systems with smaller average currents, which stems from more equal positive and negative portions of the charge-current density.
For supporting details, we refer to Appendix~\ref{app:current_density}.

Apart from the clear trend outlined above in terms of the near-edge current behavior, we note two outlier behaviors.
First, for sufficiently small system sizes, there is an increasing amount of edge-edge contributions with perpendicular relative orientation also in systems with high rotation symmetry $C_n$, which consequently causes a negative near-edge charge-current density.
If these contributions obtain a larger weight than the more parallel contributions in the Fermi surface average, then this negative contribution can even induce a sign change in the average current with shrinking system size, see Sec.~\ref{sec:results:currents:magnetic_signatures}.
Second, the triangular and square systems are exceptions. The triangular system is the only concave geometry, while the square system shows a highly destructive interference resulting in aggravated suppression of the order parameter, as we later discuss in Sec.~\ref{sec:results:phase_diagrams}.
These effects lead to notably different edge-edge contributions compared to the other geometries, yielding small negative charge-current density for the square, while it yields a small positive charge-current density for the triangle.

Finally, we briefly comment on the behavior at higher temperatures $T$.
In a semi-infinite system, increasing temperature generally leads to a monotonic reduction of the charge-current density, due to the superconducting order parameter $\Delta(T)$ being reduced relative to the zero-temperature bulk value $\Delta_0$.
However, we find that there are slightly non-monotonic effects with increasing temperature in finite-sized systems since also the system size $\mathcal{D}/\xi(T)$ effectively shrinks due to an increasing coherence length $\xi(T)/\xi_0 \approx \Delta_0/\Delta(T)$, which in turn can enhance the current through stronger edge-edge effects. For details, we refer to Appendix~\ref{app:current_density_temp}.

\subsection{Net current and magnetic signatures}
\label{sec:results:currents:magnetic_signatures}
\begin{figure}[t!]
	\includegraphics[width=\columnwidth]{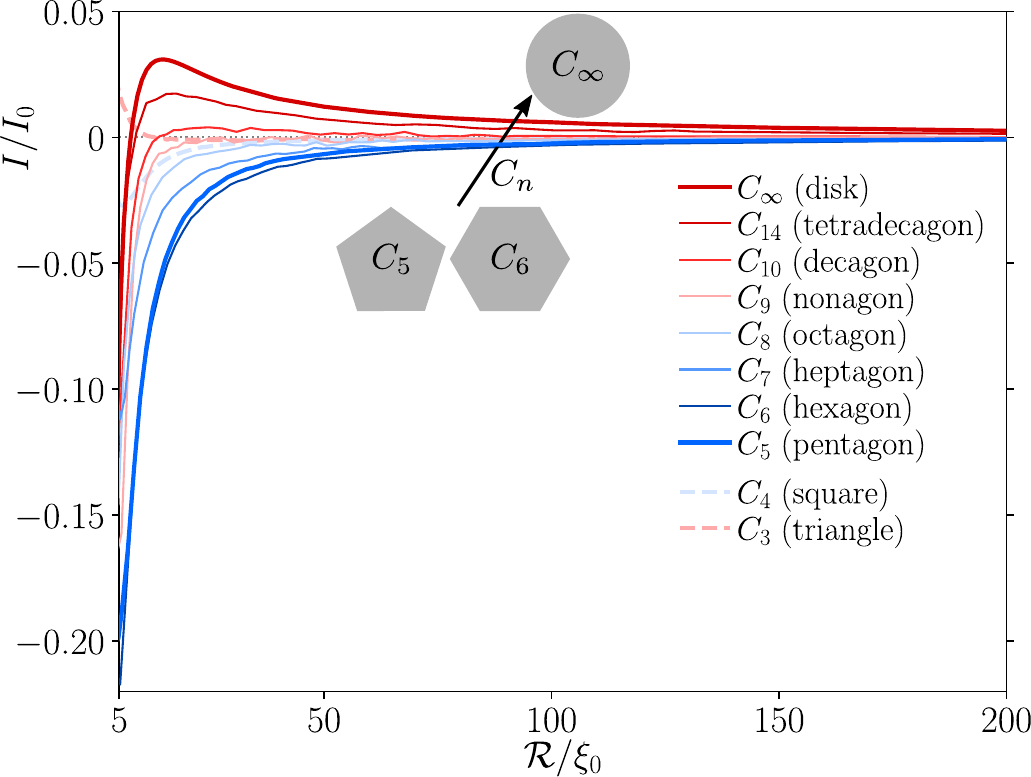}
	\caption{Net current $I$ [Eq.~(\ref{eq:net_current})], integrated from the edge to the system center, as a function of system radius $\mathcal{R}\sim\mathcal{D}/2$ for systems of different geometric shapes (line colors), with negative chirality $\hat{\nu} = -\hat{z}$, temperature $T=0.1\Tc$, and natural units $I_0 \equiv j_0\xi_0 = \hbar|e|\NF\vF^2$. Dotted line marks $I=0$.}
	\label{fig:net_current}
\end{figure}

\begin{figure}[t!]
	\includegraphics[width=\columnwidth]{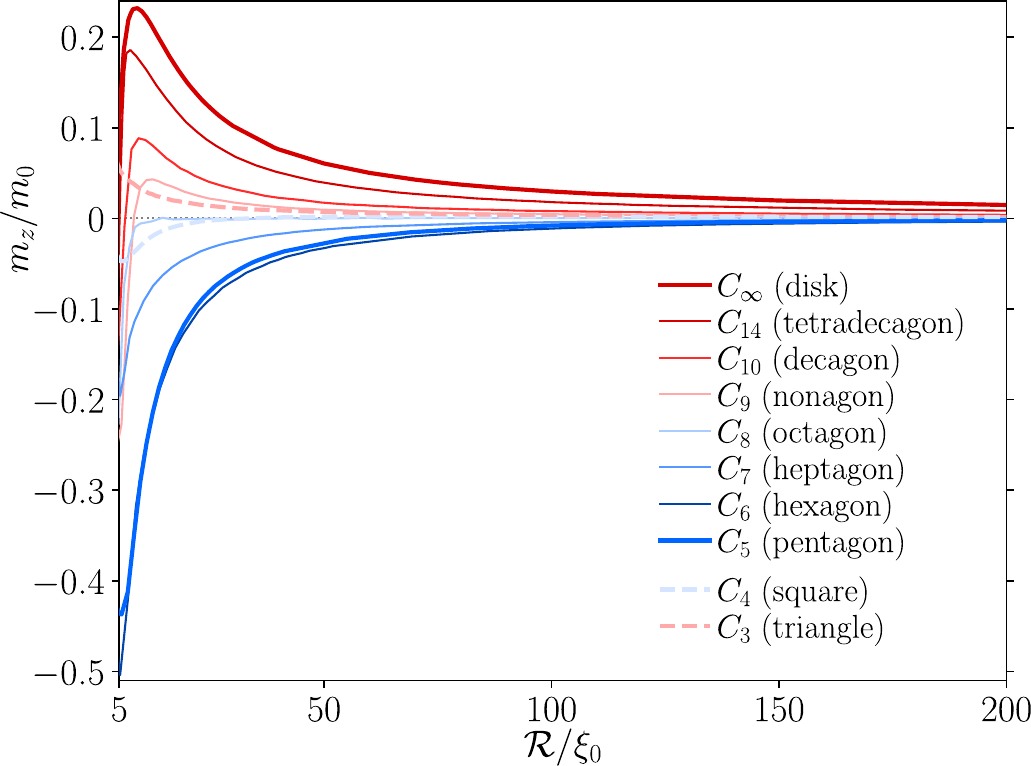}
	\caption{Same as Fig.~\ref{fig:net_current} but for the total area-averaged OMM $m_z$ [Eq.~(\ref{eq:model:magnetic_moment})], with units $m_0 \equiv (N/2)\muB$, total number of Cooper pairs  $N/2$, and Bohr magneton $\muB \equiv \hbar|e|/2m^*$.}
	\label{fig:orbital_magnetic_moment}
\end{figure}

\begin{figure}[t!]
	\includegraphics[width=\columnwidth]{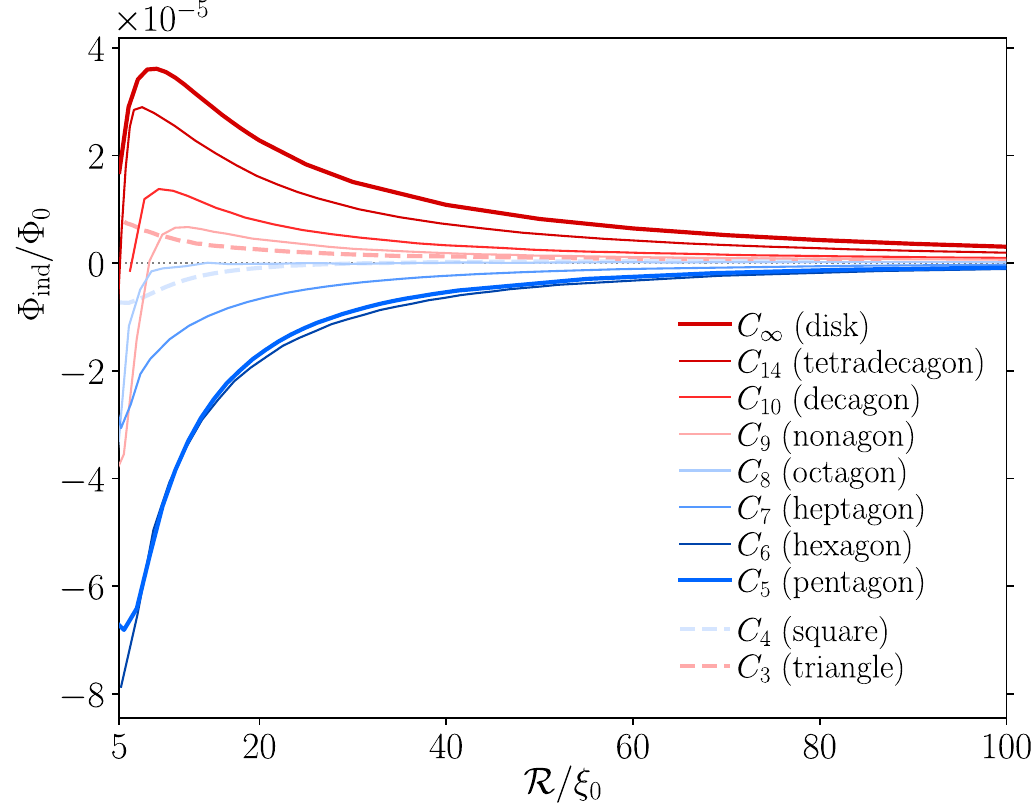}
	\caption{Same as Fig.~\ref{fig:net_current} but for the total induced magnetic flux $\Phiind$ [Eq.~(\ref{eq:model:total_flux})] in units of the flux quantum $\Phi_0\equiv hc/2|e|$, at $\lambda_0 = 40\xi_0$ (all other quantities are computed for $\lambda_0\to\infty$, since they vary minimally with $\lambda_0$ in this regime, see Sec.~\ref{sec:methods:model}).}
	\label{fig:induced flux}
\end{figure}

Having established the spatial dependence of the charge-current density $\vj(\vR)$, we next focus on the total, or integrated, net current, and also the associated magnetic signatures.
Again, we are able to demonstrate an equivalent systematic evolution with shape and size.
We compute the net current running parallel to the edge (i.e.~the full sheet current) as 
\begin{equation}
\label{eq:net_current}
    I = \int_0^{\mathcal{R}} d\rho j_\phi(\rho),
\end{equation}
which has the same discrete (continuous) rotation symmetry $C_n$ around the system center as the polygon (disk) system itself.
For the polygons, we choose $\phi$ such that we integrate from the middle of an edge to the system center, which is a representative choice (see comment at the end of this section).
We plot $I$ in Fig.~\ref{fig:net_current} as a function of system radius $\mathcal{R} \sim \mathcal{D}/2$ for many different regular system shapes (line colors).
We find that $I$ evolves smoothly with system size $\mathcal{R}$ for each shape with clear trends when changing the shapes.
We discuss these results together with the fully area-averaged OMM $m_z$ and total flux $\Phiind$ plotted in Figs.~\ref{fig:orbital_magnetic_moment} and \ref{fig:induced flux}, respectively, since they show the same overall scaling behavior.

First of all, we find that all systems show a vanishing net current and magnetic signatures in the limit $\mathcal{R}\to\infty$, with the disk having the slowest asymptotic decay.
This is a direct consequence of edge-edge contributions continue to be important also for rather large disks, since the disk always has a finite edge curvature.
In large finite systems we also find a clear smooth behavior from negative to positive currents, OMM, and magnetic flux with increased rotation symmetry from $C_5$ to $C_\infty$, see arrow in Fig.~\ref{fig:net_current} (the triangle and square shapes are outliers as explained in Sec.~\ref{sec:results:currents:current_density}).
Second, we find that all finite-sized systems show drastically increased net current, and associated magnetic signatures when the system size is reduced, due to increasing finite-size effects, specifically the edge-edge contributions to the surface propagator.
In fact, we find that the net current and magnetic signatures even diverge for smaller system size as~$I \sim 1/\mathcal{R}$, as also found in Ref.~\cite{Holmvall:2023:enhanced}.
However, this divergence is cut-off for sufficiently small system sizes due to the strong suppression of the order parameter, which we investigate in Sec.~\ref{sec:results:phase_diagrams}.
This leads to local maxima for the magnitude of the current at $\mathcal{R} \sim 5\text{--}10\xi_0$, with especially the pentagon and hexagon hosting large negative currents, while the disk also have large but instead positive currents, and equivalent for the magnetic signatures.
We further note that systems with high rotation symmetry can show a sign change in the net current and magnetic signatures at sufficiently small $\mathcal{R}$.
This is related to an increasing amount of edge-edge contributions between perpendicular relative edge orientations as explained in Sec.~\ref{sec:results:currents:current_density}.

Most interestingly, despite the cut-off in currents due to suppressed superconductivity in very small systems, we find that all systems show a dramatic enhancement of the net current and magnetic signatures in finite-sized systems, $\mathcal{R} < 50 \xi_0$.
For instance, the net current in Fig.~\ref{fig:net_current} even becomes comparable with the large values typically obtained in chiral $p$-wave superconductors~\cite{Wang:2018}.
Furthermore, the magnitude of the OMM in Fig.~\ref{fig:orbital_magnetic_moment} becomes as large as $0.5(N/2)\muB$, i.e.~$50\%$ of the Bohr magneton per Cooper pair, which is comparable with the maximal OAM $(N/2)\hbar$ reported in chiral $p$-wave systems~\cite{Sauls:2011} (the OMM is the charged analogue of the OAM).
In addition, we find that the maximum total flux (per 2D layer) is roughly $8 \times 10^{-5}\Phi_0$, see Fig.~\ref{fig:induced flux}.
Considering a system with $\mathcal{R}\sim5\text{--}10\xi_0$ and a coherence length on the order of a $2\text{--}10$ nanometers, this yields a magnetic field on the order of $0.01\text{--}0.5$~mT.
These results should be within the measurable range of state-of-of-the-art experiments~\cite{Persky:2022}, using e.g.~scanning probes and cantilever setups as schematically illustrated in Fig.~\ref{fig:sketch_setup}.

Finally, if we consider also increasing temperatures, we find that it generally leads to an overall suppression of the above quantities, but with slightly non-monotonic behavior.
Specifically, changing temperature effectively changes the system size $\mathcal{D}/\xi(T)$, which modifies the strength of the finite-size effects in small systems, while in larger systems the current can approach a very small finite value instead of zero at intermediate temperatures, also consistent with previous results~\cite{Wang:2018}.
Moreover, we note that plotting the above quantities as functions of e.g.~side length, circumference or area does not qualitatively change the results or interpretations.
We further note that the definition of the net current in Eq.~(\ref{eq:net_current}) is representative since it has the same scaling behavior as the fully area-averaged magnetic signatures, which is not necessarily obvious since they e.g.~have different weighing with radial coordinate in the integrand and magnetic screening~\cite{Holmvall:2023:enhanced}, respectively, see Eqs.~(\ref{eq:model:magnetic_moment})--(\ref{eq:model:total_flux}) vs Eq.~(\ref{eq:net_current}).
For instance, $\Phiind$ scales effectively as $(\xi_0/\lambda_0)^2$ and may thus become larger with smaller penetration depth $\lambda_0$, while both $I$ and $m_z$ instead become smaller~\cite{Holmvall:2023:enhanced}.

In summary, our results highlight mesoscopic patterning into specific mesoscopic shapes as a highly viable route to produce finite spontaneous currents and associated magnetic signatures, which can be used to verify chiral superconductivity using a number of different experimental techniques~\cite{Geim:1997,Chibotaru:2000,Kanda:2004,Grigorieva:2006,Kokubo:2010,Cren:2011,Gustafsson:2013,Timmermans:2016,Curran:2023}.
We especially find pentagons, hexagons, and disks as the most promising geometries as they host the strongest current and magnetic signals, while the square and triangle shaped samples hosts the smallest signatures. 

\section{Geometric order parameter suppression and competing orders}
\label{sec:results:phase_diagrams}
In the previous Sec.~\ref{sec:results:currents} we showed how the chiral currents and associated magnetic signatures are dramatically enhanced towards smaller system sizes.
However, this enhancement is eventually cut-off at the smallest system sizes. Here we explain this cut-off behavior by showing how 2D confinement leads to a suppression of the chiral $d$-wave state due to a suppression of the underlying nodal $d$-wave components. This in turn  triggers a competition both between different pairing symmetries and the normal state, which we here quantify via the ground-state phase diagram.
We here focus on the square and disk geometries since these shapes reflect the extreme cases where either only one or both nodal components are suppressed, respectively.
To complement these results, we report on other geometries and geometric effects in Appendix~\ref{app:order_parameter}.

In Fig.~\ref{fig:nodal_suppression} we illustrate the spatial dependence of the two nodal $d$-wave order parameter components of a chiral $d$-wave superconductor in a small square and disk.
For the square geometry in Figs.~\ref{fig:nodal_suppression}(a)--\ref{fig:nodal_suppression}(c), the edges are aligned with the crystal $ab$-axes and thus aligned with the order parameter lobes of the $\Dxtyt$ component (solid) and nodes of the $\Dxy$ component (dashed), respectively, see Fig.~\ref{fig:basis_functions} for reference.
The square edges are therefore pair-breaking for the $\Dxy$ component~\cite{Hu:1994,Jian:1994,Lofwander:2001,Vorontsov:2018}.
Specifically, a quasiparticle impinging on an edge scatters between different signs of the $\Dxy$ component, thus accumulating an effective $\pi$ phase shift.
Such a process is related to a Jackiw-Rebbi zero mode~\cite{Jackiw_Rebbi:1976}, and consequently
the emergence of fermionic bound states, which suppress the $\Dxy$ component at the edge~\cite{Sauls:2011}, as seen in Figs.~\ref{fig:nodal_suppression}(b) and \ref{fig:nodal_suppression}(c), while the $\Dxtyt$ component is slightly enhanced at the edges as a reaction.
We note that the enhancement of the $\Dxtyt$ component is most notable at low temperature, in contrast to recently proposed boundary-enhanced superconductivity at $\Tc$~\cite{Samoilenka:2020,Samoilenka:2021,Hainzl:2022,Roos:2025,Roos:2025:b}.
Henceforth, we focus on the order parameter suppression as that sets the behavior, while the bound states have already been extensively discussed, see e.g.~Refs.~\cite{Sauls:2011,Holmvall:2023:enhanced}.

\begin{figure}[t!]
	\includegraphics[width=\columnwidth]{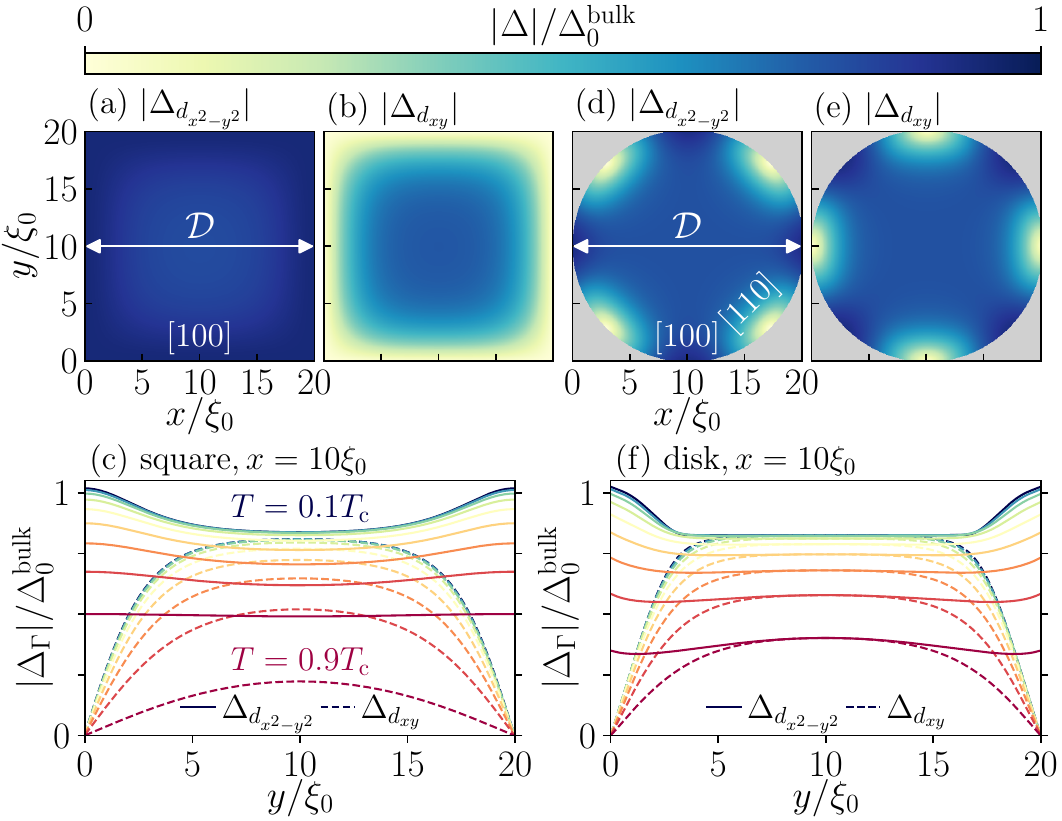}
	\caption{(a),(b) Magnitude of the order parameter components $\Dxtyt$ and $\Dxy$, respectively, in a chiral $d$-wave superconductor with negative chirality $\hat{\nu}=-\hat{z}$ and square shape of size $\mathcal{D}=20\xi_0$ with edges aligned with the crystal $ab$-axes at temperature $T=0.5\Tc$. (c) Magnitudes across the square center ($x=\mathcal{D}/2$) at different temperatures (line colors). (d)--(f) Same as (a)--(c) but in a disk-shaped superconductor. Here, $\Delta_0^{\rm bulk} \approx 1.51\kB\Tc$ is the bulk gap at $T=0$ in a single-component nodal $d$-wave superconductor.}
	\label{fig:nodal_suppression}
\end{figure}

In contrast, for the disk system in Figs.~\ref{fig:nodal_suppression}(d)--\ref{fig:nodal_suppression}(f), the edge orientation changes continuously along the perimeter with respect to the crystal $ab$-axes.
There is thus a local, but on average equal, suppression (enhancement) along the nodes (lobes) of both components along the circumference of the system, see Figs.~\ref{fig:nodal_suppression}(d)--\ref{fig:nodal_suppression}(e), which are equivalent under a $45^{\circ}$ rotation.
As a consequence, the disk system on average maintains the degeneracy between the nodal components in the system interior due to an overall equal suppression. This is in contrast to the square system where the degeneracy is broken throughout the sample, see Fig.~\ref{fig:nodal_suppression}(c).
Figures \ref{fig:nodal_suppression}(c) and \ref{fig:nodal_suppression}(f) also illustrate that the suppression and degeneracy breaking is stronger at elevated temperatures. This is partly related to an overall thermal suppression of the order parameter $\Delta(T)$.
Elevated temperatures also increase the effective coherence length $\xi(T)/\xi_0 \approx \Delta_0/\Delta(T)$, such that the system size $\mathcal{D}/\xi(T)$ effectively shrinks~\cite{Holmvall:2023:enhanced}, which in turn increases edge-edge interference, further suppressing the order parameter.
This edge-edge hybridization becomes especially pronounced when the system size $\mathcal{D}$ is comparable with the order parameter healing length $r_0$, i.e.~such that the order parameter suppression at two opposite edges overlaps.
These dual reasons for order parameter suppression are for instance seen by comparing the temperature dependence of the $\Dxtyt$ component in the square in Fig.~\ref{fig:nodal_suppression}(c), which mainly shows a suppression due to the thermal excitations, against the other component or disk system in Fig.~\ref{fig:nodal_suppression}(f) where the edge-edge interference also comes into play.

\begin{figure}[t!]
	\includegraphics[width=\columnwidth]{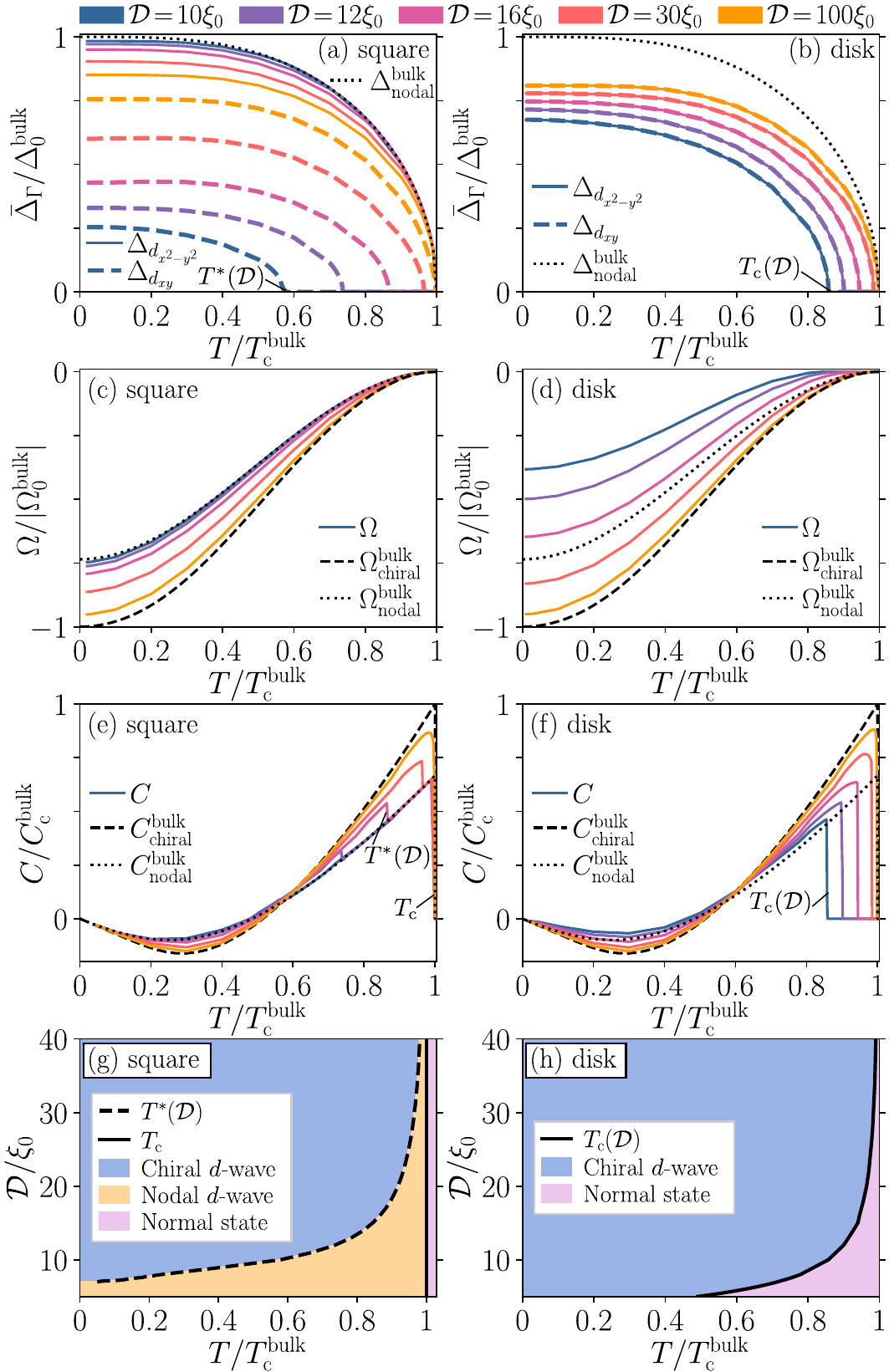}
	\caption{(a,b) Area-averaged magnitude of the order parameter components $\Dxtyt$ (solid) and $\Dxy$ (dashed) as a function of temperature $T$ for different system sizes $\mathcal{D}$ (colors) in a chiral $d$-wave superconductor with negative chirality shaped like a square and disk, respectively. Dotted black line shows the same but for a bulk nodal superconductor. (c,d) Total free energy $\Omega=\Omega_{\rm S} - \Omega_{\rm N}$ and (e,f) heat capacity $C=C_{\rm S} - C_{\rm N}$ of the superconducting (S) ground state relative to the normal state (N) as a function of temperature $T$ for the corresponding systems in (a,b) (solid). Dashed and dotted lines are for bulk chiral and nodal $d$-wave superconductors, respectively. The BCS limits of the bulk chiral $d$-wave superconductor are $\Omegabulk \approx -1.56 \mathcal{A}\NF(\kB\Tc)^2$ at $T=0$ and $\Cbulk \approx 9.38\mathcal{A}\NF\kB^2\Tc$ at $T=\Tc$~\cite{Holmvall:2023:enhanced}.
    (g,h) Ground-state phase diagram as a function of temperature and system size, in the square and disk, respectively. }
	\label{fig:op_free_energy_T}
\end{figure}

To further quantify the order parameter suppression in small systems, we plot in Figs.~\ref{fig:op_free_energy_T}(a) and \ref{fig:op_free_energy_T}(b) the area-averaged order parameter magnitudes $\bar{\Delta}_\Gamma \equiv \int dR |\Delta_\Gamma(\vR)|/\mathcal{A}$ as a function of temperature for the square and disk, respectively. 
These results illustrate that both nodal order parameter components are gradually suppressed with either higher temperature or smaller system size, eventually completely vanishing.
Importantly, however, the order parameter suppression is on average unequal (equal) in the square (disk) system between the two nodal $d$-wave components, thus breaking (maintaining) the overall degeneracy.
In particular, for the square system the $\Dxy$ component is completely suppressed at some $T^*(\mathcal{D}) < \Tcbulk$ for small systems, due to its strong suppression at each edge of the square.
Still, the same system remains superconducting due to a finite $\Dxtyt$ component until the regular thermal suppression at $\Tcbulk$, as also confirmed by the negative free energy $\Omega<0$ with respect to the normal state for all temperatures in Fig.~\ref{fig:op_free_energy_T}(c).
Interestingly, the broken degeneracy between the two $d$-wave components is manifested even in relatively large square systems up to $\mathcal{D}=100\xi_0$ as seen in Fig.~\ref{fig:op_free_energy_T}(a).
In contrast, in the disk system, both nodal $d$-wave components are instead suppressed equally, causing a vanishing of the superconducting state at $T_c(\mathcal{D}) < \Tcbulk$ for small systems, as also confirmed by the free energy $\Omega$ becoming positive at the same temperature in Fig.~\ref{fig:op_free_energy_T}(d).
The disk system converges particularly slow towards the bulk value, similar to the slow asymptotic $\sim 1/\mathcal{D}$ behavior found in the charge current towards the semi-infinite limit in Sec.~\ref{sec:results:currents}.
Taken together, we find that confinement induces a competition between the chiral and nodal superconducting states and the normal state, set by the system geometry.

To fully quantify the competition between different states, we plot the heat capacity in Figs.~\ref{fig:op_free_energy_T}(e) and \ref{fig:op_free_energy_T}(f), and the ground-state phase diagram as a function of temperature and system size in Figs.~\ref{fig:op_free_energy_T}(g) and \ref{fig:op_free_energy_T}(h).
Here, $T^*(\mathcal{D})$ is the size-dependent transition between the chiral and nodal superconducting states, also marked in Fig.~\ref{fig:op_free_energy_T}(a), while $\Tc(\mathcal{D})$ is the size-dependent transition to the normal state as also marked in Fig.~\ref{fig:op_free_energy_T}(b).
We find that the temperature suppression of the order parameter components are smooth, leading to a second-order phase transition with a jump in the heat capacity.
The jump at the normal-superconducting phase transition at $\Tc$ is comparable to that in a bulk systems, denoted $\Cbulk$, in both geometries in Figs.~\ref{fig:op_free_energy_T}(e) and \ref{fig:op_free_energy_T}(f).
Interestingly, there is an additional phase transition and jump at $T^*(\mathcal{D})$ in the square system in Figs.~\ref{fig:op_free_energy_T}(e) and \ref{fig:op_free_energy_T}(g), which occurs more generally in any system that breaks the degeneracy of the underlying nodal components.
This second jump is proportional to the difference in heat capacity between the gapless nodal state and the fully gapped chiral state, which have polynomial and exponential temperature dependence $C(T)$, respectively.
The jump therefore grows with increasing temperature and can become larger than $10\%$ of the bulk jump $\Cbulk$ as shown in Fig.~\ref{fig:op_free_energy_T}(e). This relatively large secondary jump in heat capacity and the crossover between polynomial and exponential $C(T)$ at $T^*(\mathcal{D})$ could therefore serve as indirect signatures of chiral superconductivity, which may be measurable using nanocalorimetry~\cite{Tagliati:2012,Willa:2017,Feng:2019}.
Next, we note that other regular shapes show a similar phase diagram to the disk system in Fig.~\ref{fig:op_free_energy_T}(h) but with a faster convergence to the semi-infinite limit, see Appendix~\ref{app:order_parameter}.
Finally, we speculate that these mesoscopic shape and finite-size effects can be further enhanced by non-magnetic impurities~\cite{Seja:2025} and Fermi surface effects~\cite{Wennerdal:2020}, especially as the Fermi velocity and effective coherence length increase.
Moreover, normal state anisotropy might lead to either stronger or weaker edge-edge interference depending on the orientation of the system edges with respect to the crystal $ab$-axes.

In summary, we show that mesoscopic finite-size effects induce a geometric suppression of the chiral $d$-wave state, caused by a suppression of the underlying nodal $d$-wave components.
We show that this suppression directly depends on the geometry of the system, inducing a shape-dependent competition with second-order phase transitions between different pairing symmetries and the normal state. The phase transitions are quantified by relatively large jumps in the heat capacity that may be measurable with nanocalorimetry~\cite{Tagliati:2012,Willa:2017,Feng:2019}.
Our results indicate that the superconducting pairing symmetry can be designed to some extent by mesoscopic patterning.

\section{Concluding remarks}
\label{sec:conclusions}
Chiral $d$-wave superconductivity has been proposed in a multitude of materials, but experimental verification remains an outstanding issue for all chiral superconductors~\cite{Bjornsson:2005,Kirtley:2007,Hicks:2010,Iguchi:2024}.
To make matters worse, theory predicts that the prototypical experimental fingerprints of chiral superconductivity, namely the chiral edge currents and their magnetic signatures, may be orders of magnitude smaller or even vanishing in generic chiral $d$-wave superconductors as compared to in chiral $p$-wave superconductors~\cite{Huang:2014,Huang:2015,Tada:2015,Volovik:2015:b,Nie:2020}, although some lattice structures may elude this suppression~\cite{Black-Schaffer:2012,Pathak:2024}.
Specifically, there is a destructive contribution between the chiral edge modes and the condensate backflow~\cite{Nie:2020}, with the charge-current density changing sign at a short distance $\sim \xi_0$ from the edge, resulting in an effective close to zero net current~\cite{Wang:2018,Holmvall:2023:enhanced}.

Motivated by a need to find scenarios that can enhance potential experimental signatures in all chiral $d$-wave superconductors, we study the influence of mesoscopic shape and finite-size effects on chiral $d$-wave superconductivity.
Interestingly, we find that these effects can be used to dramatically enhance and even design the chiral charge currents and their associated magnetic signatures.
Motivated by existing experimental techniques for patterning~\cite{Geim:1997,Chibotaru:2000,Kanda:2004,Grigorieva:2006,Kokubo:2010,Cren:2011,Gustafsson:2013,Timmermans:2016,Curran:2023}, we focus primarily on mesoscopic superconductors shaped like regular polygons or a circular disk.
We find that mesoscopic finite-size effects lead to a relative enhancement of both the near-edge or far-edge portions of the charge-current density, which strongly depends on the rotational symmetry $C_n$ of the system.
Specifically, for a system with negative chirality $\hat{\nu} = -\hat{z}$, we find a clear trend of large negative net charge current and magnetic signatures in systems with low rotation symmetry, especially for pentagons ($C_5$) and hexagons ($C_6$), but which then evolves to become a large and positive net current in the circular disk limit ($C_\infty$).
In contrast, the square and triangle geometries host the smallest net currents.
We associate this shape-dependence to edge-edge interactions, mainly occurring between edges with more perpendicular (parallel) relative orientation in systems with low (high) $C_n$, which give rise to a destructive (constructive) interference effect for the current relative to the current generated by the innate dispersion of the edge states, resulting in a negative (positive) near-edge current.

Overall, we find that the net current and magnetic signatures are larger in smaller systems.
The net current may even become comparable to that in semi-infinite chiral $p$-wave superconductors~\cite{Wang:2018}, with a large magnetic moment of $\muB/2$ per Cooper pair and magnetic fields estimated to the order of $0.01\text{--}0.5$~mT.
These magnetic signatures should fall within the measurable range for state-of-the-art scanning probes~\cite{Persky:2022}.
Below these maxima the enhancement is eventually cut-off due to a finite-size suppression of the superconducting order parameter in the chiral $d$-wave state.
We even find a significant competition between the chiral and nodal superconducting states, as well as the normal state in small systems.
We quantify this competition via the full phase diagram as a function of system shape, size, and temperature.
Beyond the usual normal-superconducting transition at $\Tc$, we find that in systems that break the symmetry between the underlying nodal components (e.g~square-shaped systems), there is an additional second-order phase transition between nodal and chiral superconductivity at $T^*(\mathcal{D})<\Tc$. This additional transition is associated with a jump in the heat capacity that can be roughly $10\%$ of the bulk jump at $\Tc$, which could serve as an indirect signature of chiral superconductivity measurable with nanocalorimetry~\cite{Tagliati:2012,Willa:2017,Feng:2019}.

In conclusion, we propose mesoscopic patterning as a highly feasible route to experimentally verify chiral $d$-wave superconductivity.
Particularly promising are superconductors shaped as~pentagons, hexagons, or disks with a size of tens of $\xi_0$, thus tens to hundreds of nanometers.
Squares, on the other hand, have the the lowest currents, but have an additional jump in the heat capacity due to a phase transition into an intermediate nodal superconducting state.
We further note that earlier works~\cite{Bouhon:2014,Etter:2014,Etter:2018,Curran:2023} have also discussed aspects of sample geometry, but in chiral $p$-wave superconductors, demonstrating how the charge-current density may change directions between different edge orientations due to effects such as e.g.~high filling and modified boundary conditions, resulting in the current and magnetic fields possibly canceling.
Notably, our results occur even in the absence of such effects and are instead caused by the microscopic edge-edge contributions in the quasiparticle propagator, which act as an interference effect present in any finite mesoscopic sample.

As a future outlook, it would be interesting to study if our clear mesoscopic shape and size effects are also present in chiral $p$-wave superconductors and superfluids~\cite{Sauls:2011}.
Other interesting outlooks include future studies considering microscopic and multi-layer effects~\cite{Fidrysiak:2023,Pathak:2024}, non-trivial Fermi surface and filling effects~\cite{Bouhon:2014}, spin degrees of freedom~\cite{Seja:2024:magnetization}, non-magnetic impurities~\cite{Seja:2025}, surface roughness~\cite{Suzuki:2017,Suzuki:2024:arxiv,Higashitani:2024} and other boundary conditions~\cite{Etter:2018}, as well as non-equilibrium or transport effects~\cite{Seja:2021,Seja:2022:thermopower,Seja:2022:current_injection}.

\acknowledgements
We thank R.~Arouca, K.~M.~Seja and M.~Fogelstr{\"o}m for valuable discussions.
We acknowledge N.~Wall-Wennerdal, T.~L{\"o}fwander, M.~Fogelstr{\"o}m, M.~H\r{a}kansson, O.~Shevtsov, and P.~Stadler for their work on SuperConga.
We acknowledge financial support from the Swedish Research Council (Vetenskapsr{\aa}det) Grant No.~2022-03963 and the European Union through the European Research Council (ERC) under the European Union’s Horizon 2020 research and innovation programme (ERC-2022-CoG, Grant agreement No.~101087096).
The computations were enabled by the Berzelius resource provided by the Knut and Alice Wallenberg Foundation at the National Supercomputer Centre.
Additional computations and data handling were enabled by resources provided by the National Academic Infrastructure for Supercomputing in Sweden (NAISS) and the Swedish National Infrastructure for Computing (SNIC) at NSC, PDC, HPC2N, and C3SE, partially funded by the Swedish Research Council through grant agreements No.~2022-06725 and No.~2018-05973.
All data is publicly available~\cite{Holmvall:2025:design:data} and was generated using the open-source framework SuperConga~\cite{SuperConga:2023}.

\appendix

\section{Charge-current density: shape and size dependence}
\label{app:current_density}
\begin{figure}[t!]
	\includegraphics[width=\columnwidth]{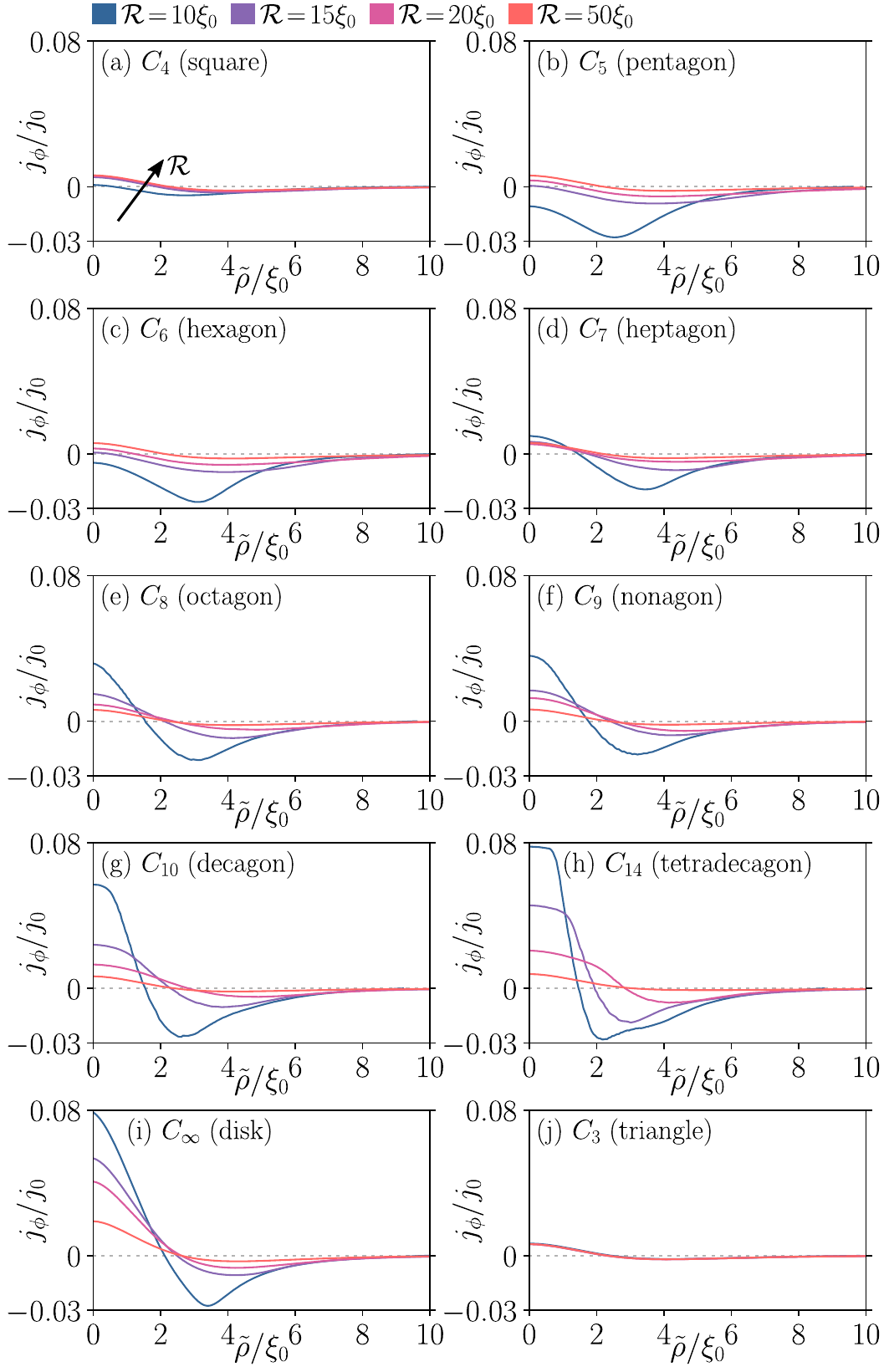}
	\caption{Azimuthal component of the charge-current density as a function of radial distance from the edge towards the center, for different system sizes (colors) and shapes (a-j), at $T=0.1\Tc$, with negative bulk chirality. Dotted lines are guides to the eye marking $j_\phi=0$.}
	\label{fig:current_density_sizes}
\end{figure}

In Sec.~\ref{sec:results:currents:current_density} we summarize shape and size dependence of the charge-current density. This Appendix contains supporting data to that discussion.
First, Fig.~\ref{fig:current_density_sizes} shows the azimuthal component of the charge current-density $j_\phi$ as a function of the radial coordinate measured from the edge, $\tilde{\rho}$, at low temperature $T=0.1\Tc$ for different system sizes (colors) and shapes (panels).
Specifically, each panel illustrates how the spatial dependence of the charge-current density varies from a relatively small system size $\mathcal{R}=10\xi_0$ (blue) to relatively large system size $\mathcal{R}=50\xi_0$ (orange).
This is the range over which the net current and magnetic signatures go from being large and either positive or negative for different shapes, to approaching the vanishing values of the semi-infinite limit, see Sec.~\ref{sec:results:currents:magnetic_signatures}.
The full spatial dependence of the charge-current density can generally be quite complicated due to both containing edge mode and condensate contributions, as well as the multiple confinement scales at play~\cite{Sauls:2011}.
To facilitate the analysis of Fig.~\ref{fig:current_density_sizes} we start by pointing out some general observations and trends, then comment on each of the geometries.

All results in Fig.~\ref{fig:current_density_sizes} show a finite charge-current density close to the edge, which decays to zero towards the bulk over the order parameter healing length $r_0\approx2\text{--}10\xi_0$~\cite{Holmvall:2023:enhanced,Sauls:2011}.
Furthermore, the charge-current density at the edge, $j_\phi(\tilde{\rho}=0)$, generally varies monotonically with system size $\mathcal{R}$ for all different shapes, see arrow in Fig.~\ref{fig:current_density_sizes}(a), as does its overall magnitude from being large at small $\mathcal{R}$ and reducing with increasing $\mathcal{R}$.
As discussed in Sec.~\ref{sec:results:currents:current_density}, we relate this enhancement at smaller $\mathcal{R}$ to the stronger mesoscopic finite-size effects.
Additionally, any sign change in $j_\phi$, if occurring, is at smaller $\tilde{\rho}$ as $\mathcal{R}$ reduces.
To explain this, we note that edge-edge hybridization has been shown to compress the chiral edge modes to a smaller region close to the edge as $\mathcal{R}$ reduces~\cite{Holmvall:2023:enhanced}. Thus, if this edge mode contribution is positive, the sign-change to a negative charge-current density moves towards the edge.
Moreover, symmetry and charge conservation enforces the charge-current density to be zero at the center of the system, which may significantly change the decay profile when the system size $\mathcal{R}$ is comparable to or smaller than $r_0$.

Next, we comment on the different shapes.
In the square-shaped system, Fig.~\ref{fig:current_density_sizes}(a), the charge-current density is overall negative for small $\mathcal{R}$ as in other systems with low rotation symmetries $C_n$, but the overall magnitude is much smaller for the square.
We find that this comes from the strong geometric suppression of the chiral state in the square system as discussed in Sec.~\ref{sec:results:phase_diagrams}, as well as a destructive edge-edge contribution that strongly reduces the density of chiral edge modes (not shown here).
Taken together, these two effects cause a stronger suppression to the charge-density than in any other geometry.
As $\mathcal{R}$ increases, we find a smooth evolution of the charge-current density as indicated by the arrow in Fig.~\ref{fig:current_density_sizes}(a).
In the pentagon and hexagon systems in Figs.~\ref{fig:current_density_sizes}(b) and \ref{fig:current_density_sizes}(c), respectively, we find an overall negative charge-current density for small $\mathcal{R}$.
As the rotation symmetry $C_n$ increases towards the disk limit, Figs.~\ref{fig:current_density_sizes}(b)--\ref{fig:current_density_sizes}(i), the near-edge charge-current density instead turns positive, smoothly increasing with rotation symmetry of the system. The farther-edge charge-current density however stays negative due to condensate backflow, causing a sign change in the charge-density profile.
Finally, for the triangle system in Fig.~\ref{fig:current_density_sizes}(j), we find a very small but overall positive charge-current density.
We attribute this to how smaller $\mathcal{R}$ yields an increase in the density of chiral edge modes that is unique for the triangle geometry (not shown here).
Increasing the system size $\mathcal{R}$ results in an overall similar behavior but with suppressed magnitudes.

\begin{figure}[t!]
	\includegraphics[width=\columnwidth]{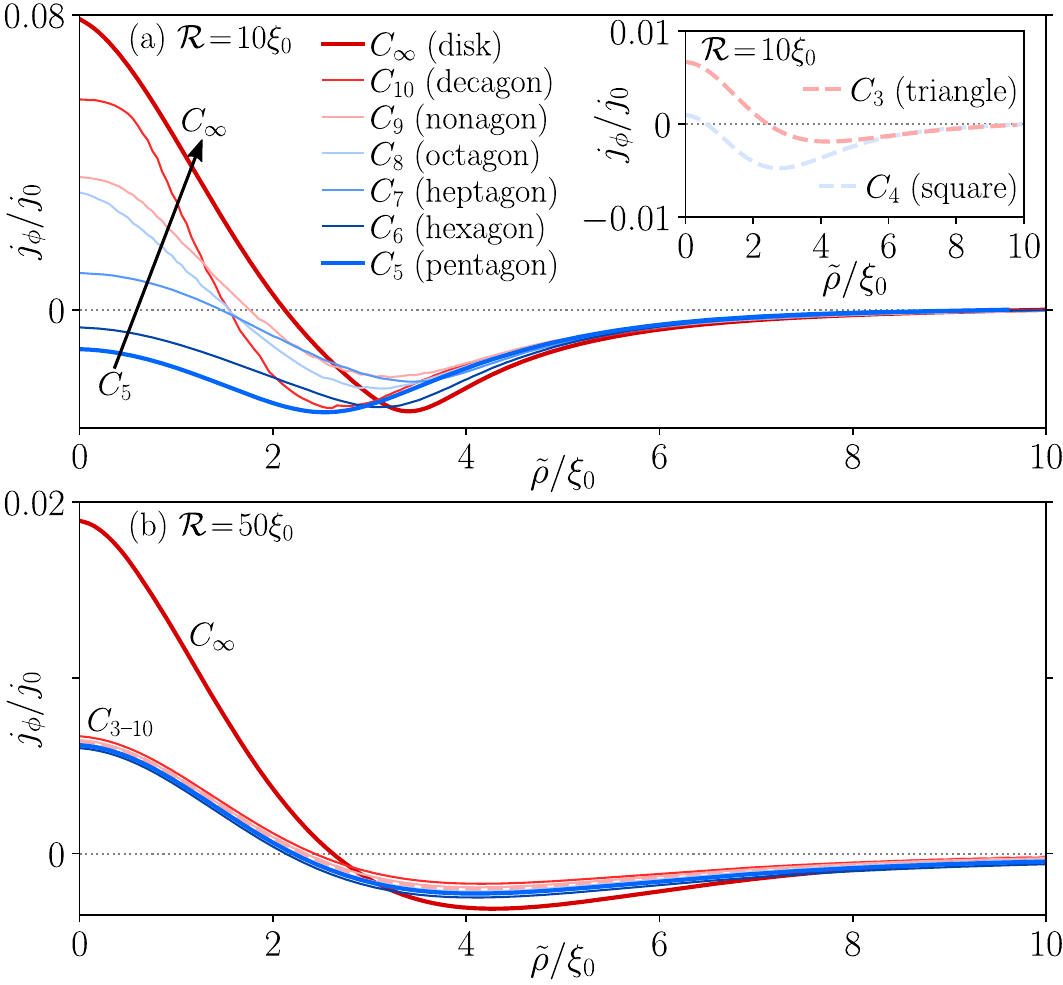}
	\caption{Same as Fig.~\ref{fig:current_density_sizes}, but with colors denoting system shape and panels denoting system size. Note the different scales on the axes. }
	\label{fig:current_density_shapes}
\end{figure}

To facilitate the analysis of how the charge-current density evolves with the rotation symmetry $C_n$ of the system, we additionally provide Fig.~\ref{fig:current_density_shapes}, where we plot $j_\phi(\tilde{\rho})$ for each shape (line colors) at fixed $\mathcal{R}$.
We start by focusing on the small system size in Fig.~\ref{fig:current_density_shapes}(a), where the portion of the charge-current density closest to the edge shows a smooth evolution from $C_5$ to $C_\infty$ as indicated by the arrow.
The outlier triangle and square geometries are plotted separately in the inset for clarity. 
As discussed in Sec.~\ref{sec:results:currents:current_density}, we attribute the smooth evolution in Fig.~\ref{fig:current_density_shapes}(a) to how the edge-edge contribution evolves with the rotation symmetry $C_n$ of the system.
To iterate, for systems with low rotation symmetry $C_n$, different portions of the edge have a more perpendicular relative orientation, resulting in a destructive edge-edge contribution to the near-edge charge-current density. For higher rotation symmetry $C_n$, this is turned to a more constructive contribution due to edges having a more parallel relative orientation. In contrast, the far-edge contribution is negative for all shapes and carried by the condensate backflow and thereby less influenced by the edge-edge contribution.
Next, Fig.~\ref{fig:current_density_shapes}(b) shows a relatively large system, $\mathcal{R}=50\xi_0$, where all polygons essentially have the same charge-current density. Hence, we find no notable shape effects.
This is easy to understand from the fact that there is minimal interference with other edges at the middle of an edge where the charge-current density is extracted, and thus minimal influence of the sample geometry.
In contrast, the disk system still shows a significantly different charge-current density, which we attribute to the finite disk edge curvature always allowing for edge-edge contribution as discussed in Sec.~\ref{sec:results:currents:current_density}.
We thus find that the disk system very slowly approaches the asymptotic values of the polygons, not fully reaching it even for $\mathcal{R}\approx200\xi_0$~\cite{Holmvall:2023:enhanced}.

\section{Charge-current density: temperature dependence}
\label{app:current_density_temp}
\begin{figure}[t!]
	\includegraphics[width=\columnwidth]{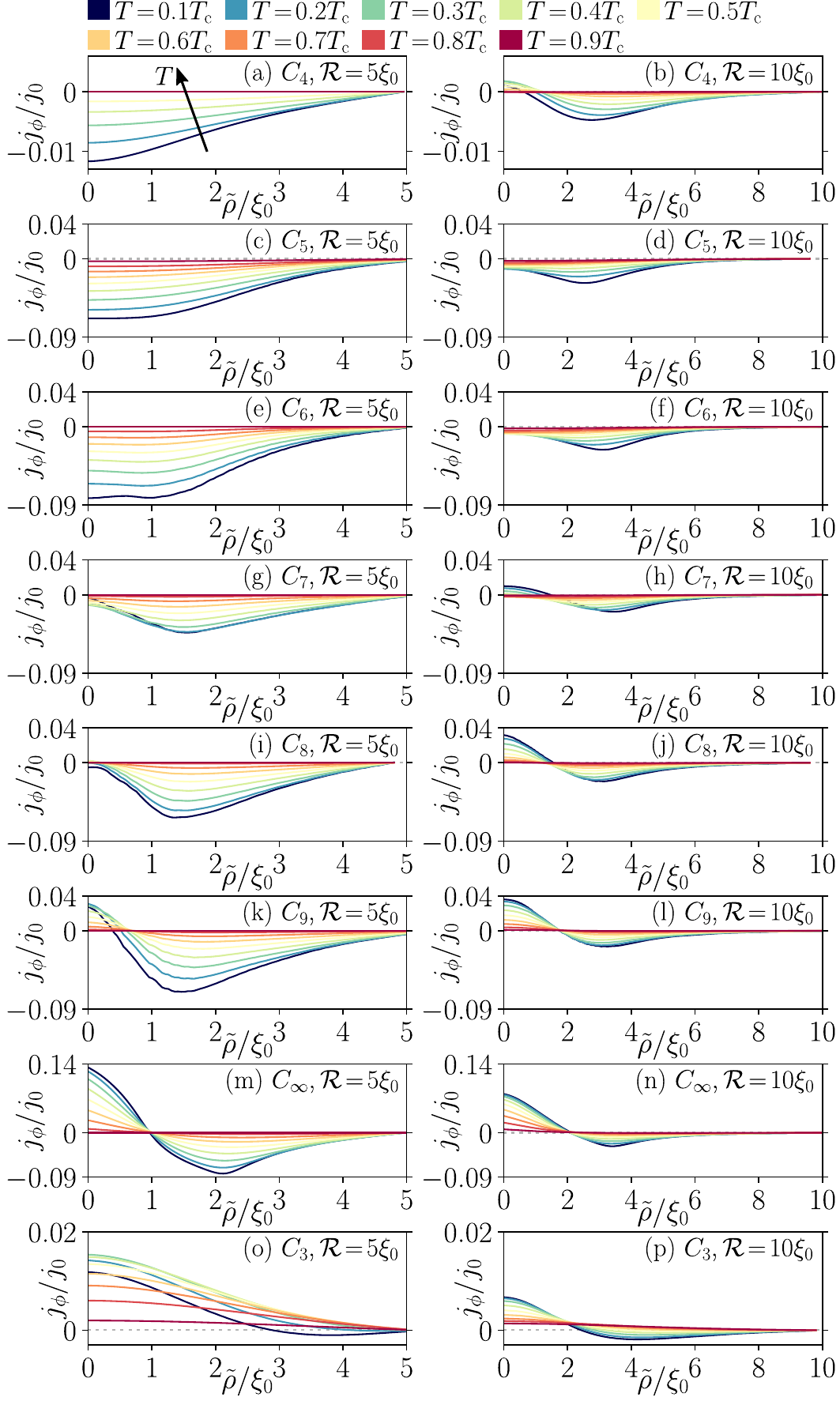}
	\caption{Same as Fig.~\ref{fig:current_density_sizes} but at different temperatures (colors) with fixed shapes and sizes (panels). Note the different scales on the axes.}
	\label{fig:current_density_temperatures}
\end{figure}

In Sec.~\ref{sec:results:currents:current_density} we summarize the temperature $T$ dependence of the charge-current density. This Appendix contains supporting data to that discussion.
In Fig.~\ref{fig:current_density_temperatures} we plot how the charge-current density varies with temperature (line colors), focusing on small system sizes (columns), since such small system sizes contain more variation between the different system geometries (rows).
In general, we find a smooth suppression of the overall charge-current density with increased temperature $T$, see arrow in Fig.~\ref{fig:current_density_temperatures}(a), which we overall relate to the thermal suppression of superconductivity $\Delta(T)/\Delta_0$.
There are, however, a few outlier results with minor non-monotonic dependence in temperature $T$, see for instance Figs.~\ref{fig:current_density_temperatures}(g,o). We attribute these results to the fact that an effectively larger coherence length $\xi(T)/\xi_0 \approx \Delta_0/\Delta(T)$ effectively also shrinks the system size $\mathcal{D}/\xi(T)$, thereby enhancing the edge-edge contributions and competing with they overall thermal suppression of superconductivity.
We find that this behavior can also occur for the other geometries at different system sizes $\mathcal{R}$.

\section{Geometric order parameter dependence}
\label{app:order_parameter}
In Sec.~\ref{sec:results:phase_diagrams} we investigate the spatial dependence of the order parameter and the finite-size suppression of the chiral $d$-wave state in superconductors shaped like squares and disks.
This Appendix contains complementary results, focusing on these effects in other sample shapes, for different alignment between system edges and crystal $ab$-axes, and for sharp vs  rounded corners.
These effects are relevant for various experimental realizations~\cite{Geim:1997,Gustafsson:2013,Timmermans:2016,Curran:2014}, and studying their influence aids the fundamental understanding of how the chiral $d$-wave state behaves and competes with other orders in mesoscopic samples.

\begin{figure}[t!]
	\includegraphics[width=\columnwidth]{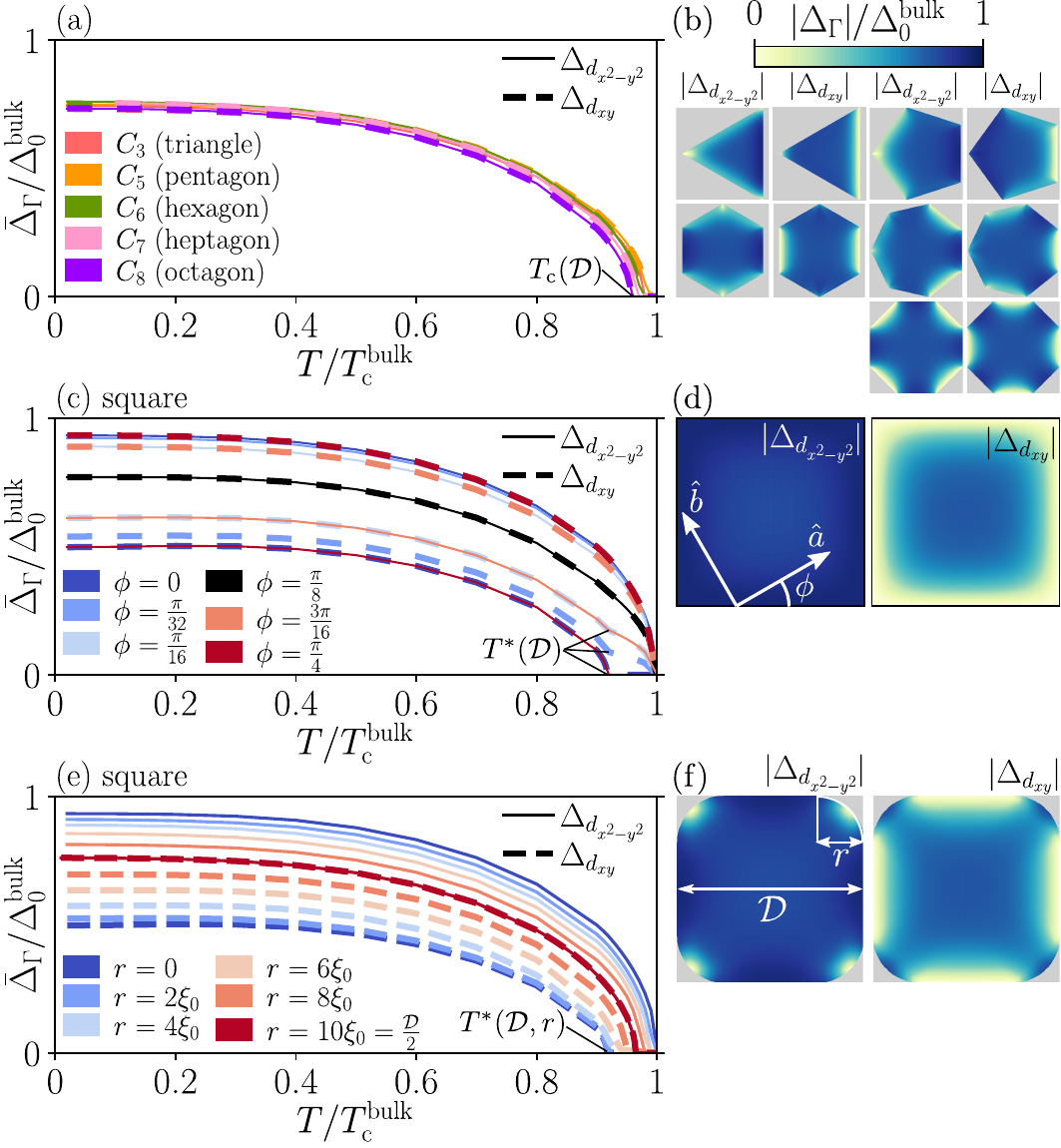}
	\caption{(a) Area-averaged magnitude of the order parameter components $\Dxtyt$ (solid) and $\Dxy$ (dashed) as a function of temperature in regular polygons of size $\mathcal{D}=20\xi_0$ for different shapes (colors) with corresponding heatmaps in (b). (c) Same as (a) but with colors denoting the angle $\phi$ between the crystal $ab$-axes in a square as indicated in (d). (e) Same as (c) but colors denoting the radius $r$ of the rounded corners of an otherwise square geometry as indicated in (f). All heatmaps are at $T=0.5\Tc$. }
	\label{fig:op_geometric_effects}
\end{figure}

In terms of the order parameter magnitudes and free energy, we find that the regular polygons beyond the square quantitatively fall close to the result of the disk system, see Figs.~\ref{fig:op_geometric_effects}(a,b), where we plot the order parameter suppression for a few representative systems.
In particular, we find that all regular polygons beyond the square on average suppress both order parameter components equally and thus behave similar to the disk system, which is explicitly illustrated by the overlapping lines in Fig.~\ref{fig:op_geometric_effects}(a).
In comparison to the disk system, however, these other geometries approach bulk behavior faster as the system size $\mathcal{D}$ grows.
We associate this difference in asymptotic behavior to the finite edge curvature only present in the disk system as $\mathcal{D} \rightarrow \infty$, which is similar to the results for the current discussed in Sec.~\ref{sec:results:currents}.
In fact, the square geometry is special since it is the only geometry where all edges only suppress one of the nodal components, since it has the same rotation symmetry $C_4$ as the nodal order parameter magnitudes.

Next, we investigate the influence of the alignment $\phi$ between the system edges and the crystal $ab$-axes in the square geometry, see Figs.~\ref{fig:op_geometric_effects}(c,d) (all other results in this work are for $\phi=0$).
This is motivated by experiment where some variation in $\phi$ may be present, depending on the superconducting sample fabrication technique~\cite{Timmermans:2016}.
We first note that there is an effective eight-fold rotation symmetry in our model (i.e.~results for $\phi=0$ and $\pi/4$ are the same), since such a rotation just exchanges the basis functions of the order parameters $\Dxtyt$ and $\Dxy$.
Importantly, for $\phi=\pi/8$, both nodal components are equally suppressed at the edges, and thus become degenerate in the sample interior, just like in the disk system.
Thus, as the angle $\phi$ increases, we find a smooth variation in the order parameter magnitudes from the square sample ($\phi=0$) towards the same behavior as in the disk system at $\phi=\pi/8$, up to some corrections due to the corners of the square (see further below).
As the angle keeps increasing from $\phi=\pi/8$ towards $\phi=\pi/4$, the order order parameter magnitudes smoothly reverts to the same results as the square system without rotation, $\phi=0$, but with $\Dxtyt$ ($\Dxy$) becoming the locally suppressed (enhanced) component at the edges instead.
Overall, we find that for a square-shaped system with generic crystal $ab$-axes rotation, the competition between nodal and chiral superconducting states falls between the results of the square with $\phi=0$ and the disk. 
In particular, as seen in Fig.~\ref{fig:op_geometric_effects}(c), $0 < \phi < \pi/4$ leads to both nodal components being finite until $\Tcbulk$, while only the $\phi=0, \pi/4$ squares are the cases where the finite geometry actually induces a different $\Tc(\mathcal{D})$ for the two order parameter components as investigated in Sec.~\ref{sec:results:phase_diagrams}. Intriguingly we find that close to $\Tcbulk$
the system may still transition from a chiral to a nodal state at some $T^{*}(\mathcal{D})$, indicated by the black lines, due to the phase difference between the two nodal components suddenly becoming zero at this point.

Next, we study the influence of rounded corners in a square geometry in Figs.~\ref{fig:op_geometric_effects}(e) and \ref{fig:op_geometric_effects}(f).
We note that Refs.~\cite{Sauls:2011} and \cite{Iniotakis:2005} have studied the influence of sharp corners on chiral $p$-wave superfluids and nodal $d$-wave superconductors, respectively, specifically showing how they give rise to double-reflection for quasiparticle states, which drastically modifies the spectrum in the vicinity of the corner.
Here, we directly quantify the influence of sharp corners on the chiral $d$-wave state by comparing the above results for the square geometry with a system where the corners have been rounded off, see Fig.~\ref{fig:op_geometric_effects}(f).
Specifically, we find that as the rounding radius $r$ increases towards $\mathcal{D}/2$, the results smoothly approach those of the disk system. Importantly, for any $r < \mathcal{D}/2$, there is still a transition from the chiral to nodal state at some temperature $T^*(\mathcal{D},r=0) < T^*(\mathcal{D},r) < \Tcbulk$ due to a suppression of the $\Dxy$ order parameter.

\bibliography{cite.bib}

\end{document}